\documentclass[published]{JHEP3} 

\JHEP{ }

\usepackage{epsfig,multicol}

\newcommand{\half}{{1\over2}}
\newcommand{\halfpi}{{1\over2\pi}}
\newcommand{\Z}{\rm Z\!\!Z}
\newcommand{\R}{\rm I\!R}
\newcommand{\I}{\rm 1\!l}
\renewcommand{\Re}{\rm Re}
\renewcommand{\Im}{\rm Im}

\newbox\pippobox

\title{Complex structure moduli stability in toroidal compactifications}

\author{Juan Garc\'{\i}a-Bellido~and Ra\'ul Rabad\'an\\
        Theory Division CERN, CH-1211 Gen\`eve 23, Switzerland\\
        E-mail: \email{bellido@mail.cern.ch}, \email{Raul.Rabadan@cern.ch}}
\received{\today}               %%
\revised{}
\accepted{}               %% These are for published papers.

\preprint{\hepth{0203247}}      
\preprint{CERN-TH/2002-067}      
\abstract{In this paper we present a classification of possible
  dynamics of closed string moduli within specific toroidal
  compactifications of Type II string theories due to the NS-NS
  tadpole terms in the reduced action. They appear as potential terms
  for the moduli when supersymmetry is broken due to the presence of
  D-branes. We particularise to specific constructions with two, four
  and six-dimensional tori, and study the stabilisation of the complex
  structure moduli at the disk level. We find that, depending on the
  cycle on the compact space where the brane is wrapped, there are
  three possible cases: i) there is a solution inside the complex
  structure moduli space, and the configuration is stable at the
  critical point, ii) the moduli fields are driven towards the
  boundary of the moduli space, iii) there is no stable solution at
  the minimum of the potential and the system decays into a set of
  branes.}


\begin{document} 

\vspace{3cm}

\section{Introduction}

Branes at angles \cite{general} provide a very rich framework for the
construction of compactifications with a chiral spectrum of a very
similar structure to the one of the standard model
\cite{standard,imr01,csu01,bklo01,cim02,bkl02}.
Generically these models are non-supersymmetric, although some
supersymmetric constructions can also be obtained \cite{csu01}. These
configurations are T-dual pictures of branes carrying non trivial
bundles wrapping the compact space \cite{b95,aads00,standard,r01}.
 
There are two types of closed string tadpoles, the
Neveu-Schwarz--Neveu-Schwarz (NS-NS) and the Ramond-Ramond (R-R)
tadpoles. The cancellation of the R-R tadpoles is a {\em necessary}
condition for the consistency of the theory. In particular, R-R
tadpole cancellation conditions guarantee the absence of chiral
anomalies in the low-energy effective theory~\cite{general,standard}.
However, even if the R-R tapdoles cancel, when supersymmetry is not
preserved the NS-NS tadpoles may appear. The system seems to be
consistent, but some potentials for the NS-NS fields are generated,
signalling that the configuration is not in a stable vacuum, and the
string vacuum has to be redefined.  This problem has been addressed in
several papers \cite{tadpoles,bkl02}.

We analyse this problem of the uncancelled NS-NS tadpoles in the
context of intersecting branes models, see also Ref.~\cite{bklo01}.
Given a cycle $\Gamma$ of a homology class on an arbitrary
compactification space, we can wrap a brane on it. The system will try
to minimise the volume of the brane, inducing a variation of the
metric moduli space. When the D-branes wrap a half homology cycle, one
can see that the potential depends {\em only} on the complex structure
moduli. Another way to see it is through the appearance of a NS-NS
tadpole term that enters in the effective action as a potential for
the complex structure moduli field. One can check that this potential
is proportional to the modulus of the periods:
\begin{equation}
|Z_{\Gamma}| = \left|\int_{\Gamma} \Omega\right| \,,
\end{equation}
where $\Omega$ is the normalised n-form in a general complex
n-dimensional manifold, and $\Gamma$ is a cycle in that class. This
form specifies the complex structure of the manifold. Sometimes,
depending on the complex structure, this brane is unstable against its
decay into other branes.

In this paper we have concentrated on tori of different (even)
dimensions, and an arbitrary number of branes. The questions we
address here are the following: given a homology class, where is the
complex structure moduli going to? Is there a minimum? Is the brane
that wraps this cycle stable at the minimum? This problem is analogous
to that studied by Moore \cite{Moore} and Denef \cite{Denef}, in their
case related to the construction of stable BPS black holes. We have
realised that the minima in both cases are exactly the same. Here we
analysed some of the results of Ref.~\cite{Moore} and extrapolated the
analysis of the minima to our case. Different phenomena can take place
in the flow of these complex structure moduli, like crossing lines of
marginal stability that make some branes decay into others
\cite{Denef}, etc.

We give here our main conclusions, and leave the description of the
details for the following sections. For the 2-dimensional torus the
complex structure moduli fields are driven to the boundary of moduli
space. In the 4-dimensional torus we find a well differentiated
behaviour depending on the wrappings of the branes around the homology
cycles. In this case, we analysed a large number of examples, although
a general description is absent, as we will discuss below. The most
interesting case, however, is the 6-dimensional torus, where we find
three different types of behaviours: i) A stable minima can be
localised in the interior of the manifold of the complex structure
moduli. This will only happen if the cycle is {\rm not} factorizable
\footnote{A 3-cycle is called factorizable if it can be decomposed
  into the product of three 1-cycles, each one wrapping a two
  dimensional torus. That is the case of most of the D-brane models
  mentioned above.}. ii) In the case there is only one factorizable
cycle, the complex structure moduli are stabilised at some points on
the boundary. iii) When the cycle can be decomposed into two
factorizable cycles, one can easily see that the minimum is at some
point in the interior of the moduli space, but the brane has decayed
into a pair of factorizable cycles. However, one can get stable
configurations in the interior of the moduli space if one considers
more than one factorizable brane. Examples of all the different types
of behaviours will be constructed. Note that we have not imposed here
the R-R tadpole cancellation conditions, although the dynamics will 
not be affected if we impose them, as we will discuss later.

When Ramond-Ramond tadpole conditions are imposed, there are, in
addition to the vacua where all branes annihilate, some specific vacua
where the non-supersymmetric sectors decouple. A trivial example where
this happens is that of a pair brane-antibrane at distant points in a
compact space. If the distance is larger than the string scale there
will be no tachyonic modes. The potential due to the NS-NS tadpoles is
proportional to the inverse of the volume of the compact space, and
arises from the sum over the winding modes. This means that the
potential will be minimised when the volume tends to infinity and the
two D-branes are very far from each other. That is what we already
know: tadpoles appear in compact spaces, but when the volume goes to
infinity its effect is like in a non-compact space. Of course, this is
only a tree-level result and quantum corrections are expected. For
example, at one-loop, there is an interaction between the pair due to
the exchange of massless string excitations. Tree-level and one-loop
interactions give two competing effects that can change the direction
of the flow.

Moreover, this uncancelled tadpoles could have very interesting
physical applications. For instance, the scalar potentials arising
from dynamical variations of internal compactification spaces (i.e.
complex structures) could be used as inflaton potentials for
cosmological inflationary scenarios from strings~\cite{inflation}.

In the following sections we will give an introduction to complex
structures and moduli spaces, in order to understand the
classification of such scalar potentials. We will also give the
necessary stability criteria that may help determine phenomenological
consequences like inflation. The outline of the paper is the
following: in Section 2 we give a general discussion of toroidal
compactifications; Sections 3, 4 and 5 discuss some remarkable cases
for the two, four and six-dimensional tori, respectively. In the last
two Sections we review the work of Moore \cite{Moore}, and give the
stability criteria around the various critical points, as well as the
general solution for the 6-dimensional torus.

\section{Discussion of the models}

\subsection{Moduli of complex structures}

Consider a the 2n-dimensional tori and define a holomorphic n-form
$\Omega_0$, written as $\Omega_0 = dz_1 \wedge ... \wedge dz_n$, where
the complex coordinates $z$ depend on real coordinates $x$ and $y$ as
$dz_i = dx_i + \tau_{ij} dy_i$. The $\tau_{ij}$ are complex numbers
that specify the complex structure of the manifold.  The flat metric
on the torus can be written in terms of the complex coordinates as
$ds^2 = dz d\bar{z}$, and the K\"ahler form is $\omega = dz \wedge
d\bar{z}$. The volume of the manifold can be written in terms of the
$\Omega_0$ form as:
\begin{equation}\label{vol}
{\rm Vol} = (-1)^{n(n-1)/2} i^n \int \Omega_0 \wedge \bar\Omega_0 \,.
\end{equation}
One can always define a normalised n-form such that its total volume
is normalised to 1, i.e. $\Omega \equiv e^{{\cal K}/2} \Omega_0$. The
volume (\ref{vol}) defines a K\"ahler potential for the complex
structure moduli, ${\cal K} = -\ln({\rm Vol})$, and an induced K\"ahler
metric, $g_{IJ} = \partial_I \partial_J {\cal K}$, which normalises the
complex structure kinetic terms,
\begin{equation} 
e^{-2 \phi} g_{IJ} \partial_{\mu}Z^I \partial^{\mu}Z^J\,,
\end{equation}
where $I,J$ are coordinates in the complex structure moduli, the
$\tau_{ij}$ for instance. These kinetics terms are obtained from the
reduction of the Hilbert-Einstein action on the particular manifold.
The dependence on the dilaton comes from the closed-string tree-level
amplitude.

\subsection{Description of the system}

We will consider only D6-branes of Type IIA string
theory,\footnote{Through T-dualities one can easily generalise to
  other Dp-branes within Type IIA string theory. However, one should
  then realise that we have to take into account both K\"ahler and
  complex structure moduli.} wrapping 3-cycles on the 6-dimensional
compact space, and expanding along the other 4-dimensional Minkowski
coordinates. The D6-brane will try to minimise its volume within the
same homology class. Depending on the point on the moduli space, the
D-brane system can be stable or unstable to the decay to other
D-branes whose sum belongs to the same class. The complex structure
moduli will vary due to the potential of the NS-NS tadpoles,
triggering different effects along their evolution. This can be
generalised to T-dual configurations of Type I theory by including
orientifold planes and the orientifold images of the branes. We will
briefly discuss this case in relation to the R-R tadpole conditions,
but we will not analyse it in detail given the huge number of objects
involved.

In this section we will review how to obtain these NS-NS tadpoles, the
R-R tadpole cancellation conditions, the evolution of complex
structure moduli due to the NS-NS tadpoles, the possible decays of
D-branes through the lines of marginal stability and a discussion
about the stability of the critical points of the potential.

\subsection{R-R Tadpoles}

In order to obtain a consistent compactification one has to impose the
cancellation of all the Ramond-Ramond tadpoles. In particular, they
guarantee that the low energy chiral spectrum is anomaly free.  These
tadpole conditions tell us that the sum of the R-R charges of all
branes must be equal to zero in the case of Type IIA
compactifications, or equal to the orientifold charge in the case of
T-dual compactifications of Type I string theory. These charges are
specified by the homology class of the cycles where the branes are
wrapped,
\begin{equation}
\sum_a \Gamma_a = 0\,,
\end{equation}   
for Type IIA theory, and 
\begin{equation}\label{tadpoleI}
\sum_a \Gamma_a = q_o \Gamma_o\,,
\end{equation}
for the dual of Type I theory, where $q_o$ is the R-R charge of the
orientifold plane and $\Gamma_o$ is the cycle where it is wrapped.
These conditions tell us that, whatever the combinations and
decays of branes, the system must have a total R-R charge equal to
zero (in the Type IIA case) or equal to $q_o \Gamma_o$ (in the
T-dual of Type I).

In this paper we will analyse configurations where the R-R tadpole
conditions are not explicitly satisfied. Only in the last part of the
paper we will comment about a way to cancel them by including other
branes and antibranes.

The idea is to study the flows of the complex structure fields in
these systems for a small set of branes, extracting some general
features, and then try to impose these constraints in a more
complicated system where the number of branes is substantialy
increased and the analysis is not as straightforward. One can always
consider adding to one of these simple models some branes, with the
charges necessary to cancel the R-R tadpoles, but which are kept as
spectators. For example, if we put a brane in a cycle we can always
put an antibrane in the same cycle\footnote{In order to separate the
  brane from the antibrane, we assume that the moduli space of special
  Lagrangians for a given homology class is not a point.} but at large
distances from the brane so that they do not develop a tachyonic mode.
The R-R tadpoles are immediately cancelled but the NS-NS are added,
giving just a factor two in the potential for the complex structure.
The difference will appear at one-loop in the open string description
(D-brane interaction), but we only consider the disk (tree-level)
term.  Higher order terms will change the structure of the minima, as
we will discuss later. Note that these conditions do not need to be
imposed if there are some non-compact coordinates transverse to the
branes, as happens in the dyonic black hole constructions of
Ref.~\cite{blackholes}.

\subsection{NS-NS Tadpoles}

Let us turn now to the more dynamical NS-NS tadpoles. These tadpoles
can be written as the volume of the cycle where the D-brane is
wrapped, divided by the squared root of the whole volume of the
manifold. This can be obtained directly by identification of the
tadpole from the cylinder amplitude. In the general case, one obtains
these terms by integration of the D-brane action in the compact space.
If the NS-NS tadpoles are not cancelled, potential terms will appear
in the effective action. 

We will consider that each D-brane is volume-minimising and that it
preserves some supersymmetry. In our particular case, this means that
the brane is wrapping a special Lagrangian manifold. Then the modulus
of the period where a BPS D-brane is living gives its volume, and the
NS-NS tadpole can be easily written as:
\begin{equation}\label{vola}
V_a(\phi,\tau) = e^{- \phi} |Z_{\Gamma_a}| = e^{-\phi} 
\left|\int_{\Gamma_a} \Omega\right|\,,
\end{equation}
where $\Gamma_a$ is the cycle on which the brane wraps.
If there is more than one BPS brane the potential becomes
\begin{equation}\label{pot}
V(\phi,\tau) = \sum_a V_a = e^{- \phi} \sum_a 
\left|\int_{\Gamma_a} \Omega\right|  \,.
\end{equation}
In the T-dual description of Type I theory one should also take into
account the contribution from the $2^{9-p}$ orientifold p-planes. Each
of these planes has a tension and a R-R charge equal to $-2^{p-4}$
times the tension and the R-R charge of the brane (counting the
orientifold images of the brane as independent). For the case of
O6-planes, there are 8 of them, with a tension and R-R charge $-4$
times the brane's. That gives, independently of the dimension of the
O-planes, a contribution to the NS-NS tadpoles \cite{bklo01}:
\begin{equation}
V(\phi,\tau) = -32 \left|\int_{\Gamma_o} \Omega\right|\,.
\end{equation}

The NS-NS tadpoles are always positive definite. This is obvious for
the Type IIA case, where the tadpoles are the sum of a set of positive
real numbers, which means that for this case the absolute minimum of
the potential will be the vacuum, a system where all the branes have
been annihilated (like in the brane-antibrane case), or when the
cycles where the D-branes are wrapped have degenerated to zero volume,
in the boundary of moduli space. Indeed, as we will see, depending on
the starting point, the system can evolve to the complete absence of
branes or towards points where the volume of the branes vanish.

For the T-dual picture of Type I theory one can easily prove that the
NS-NS tadpoles, 
\begin{equation}
V(\phi,\tau) = \sum_a V_a = e^{- \phi} \left[\sum_a 
\left|\int_{\Gamma_a} \Omega\right|  - 
q_o \left|\int_{\Gamma_o} \Omega\right|\right]\,,
\end{equation}
are always positive definite, by using the triangle inequality and the
R-R tadpole conditions (\ref{tadpoleI}). This means that an absolute
minimum of this configuration will occur when the periods of the
branes have the same phases and the same charges as those of the
orientifold plane, i.e. all the branes will try to be parallel to the
orientifold plane~Ref.~\cite{bklo01}.  The system will be
supersymmetric in this case.  Of course, another possibility,
analogous to the one in the Type II case, is that in which the branes
evolve to a system where some cycles can degenerate, or more
complicated possibilities if bound states are considered. In this
paper we will not analyse configurations with orientifold planes,
but is definitely worth studying the extrapolation of our analysis
to that case.

\subsection{The evolution of the complex structure moduli fields}

Here we will discuss the dynamics of the moduli fields. From the
point of view of the effective four dimensional theory, the action for
the complex structure moduli fields is of the form
\begin{equation}\label{modlag}
{\cal L}_4 = e^{-2 \phi} g_{IJ} \partial_{\mu}Z^I \partial^{\mu}Z^J\ -
V(\phi,Z^I) \,.
\end{equation}
This effective action has been obtained by dimensional reduction of
the 10-dimensional one, where the total volume factors have been
absorved in the redefinitions of the fields. From this action, we will
see that the complex structure moduli $Z^I$ will evolve towards some
critical points of the moduli space, which we will characterise below.

Since the variations of the complex structure moduli fields are
area-preserving, the Planck constant in the Hilbert-Einstein term does
not vary, and therefore the analysis of the stability of the critical
points of these potentials $V(\phi,Z^I)$ can be done in the $Z^I$,
i.e. the $\tau_{ij}$ coordinates. The stability criteria will not
change under the redefinition of $Z^I$, needed for obtaining
canonically normalised kinetic terms in the Lagrangian (\ref{modlag}),
only the speed of approach to the critical points. Therefore, in all
the figures below, we have drawn the potential $V(\phi,\tau)$ in the
$\tau_{ij}$ coordinates. We have also assumed that the dilaton is
fixed. Note that, when correctly normalised, the tree-level potential
will be proportional to the string coupling constant, $g_s=e^\phi$.
Now, since this potential is always positive, the dilaton will evolve
towards weak coupling.

\subsection{Lines of marginal stability}\label{marginal}

Two branes that are intersecting can have tachyonic modes in the
spectrum of open string excitations between them. The presence of
tachyons is related to the possibility of the decay of the system to
another one with the same charges but with a lower volume. Locally
these intersecting branes can be seen as two planes. Depending on the
dimension of these planes, they can minimise their
area~\cite{angle,d99}. Something similar happens in general
Calabi-Yau's where branes can decay to more stable systems by changing
their complex structure~\cite{km99}. This is a geometrical condition
known as the angle criterion~\cite{angle}, that coincides with the
computation of the lowest string mode in the NS-NS sector. For every
pair of branes intersecting at a point one can define some angles
following the procedure given in Refs.~\cite{hl82,angle}. This
procedure gives m angles for m-dimensional planes intersecting at a
point in a 2m-dimensional space. These angles are called
characteristic angles.
 
We will briefly describe here the six-dimensional toroidal case for
factorizable branes. See also Refs.~\cite{imr01,r01}. There are 4 scalar
fields that can become tachyonic, with masses,
\begin{equation}\label{tachyons}
\begin{array}{rl}
\alpha' m^2_1 &= \halfpi(- \theta_1 +\theta_2 +\theta_3)\,, \\
\alpha' m^2_2 &= \halfpi( \theta_1 -\theta_2 +\theta_3)\,, \\
\alpha' m^2_3 &= \halfpi( \theta_1 +\theta_2 -\theta_3)\,, \\
\alpha' m^2_4 &= 1 - \halfpi( \theta_1 +\theta_2 +\theta_3)\,,
\end{array}
\end{equation}
where the angles $\theta_i \in [0,\pi]$. The above masses are related
to stability conditions for the pair of branes. If there is no tachyon
(notice that only one of the scalar fields can be tachyonic at a time)
the two brane system is stable, made of two BPS branes breaking all
the supersymmetries. If one of the scalars becomes massless then the
system becomes supersymmetric, with the number of supersymmetries
related to the number of scalar fields that become massless at the
same time. These conditions can be represented in a three dimensional
figure in the angle space.\footnote{See the discussion and figures of
Ref.~\cite{imr01,r01}.} The conditions bound a tetrahedron where the
different regions are split into:\\[-1mm]

{\renewcommand\belowcaptionskip{1.7em} 
\renewcommand\abovecaptionskip{-.5ex} 
\EPSFIGURE{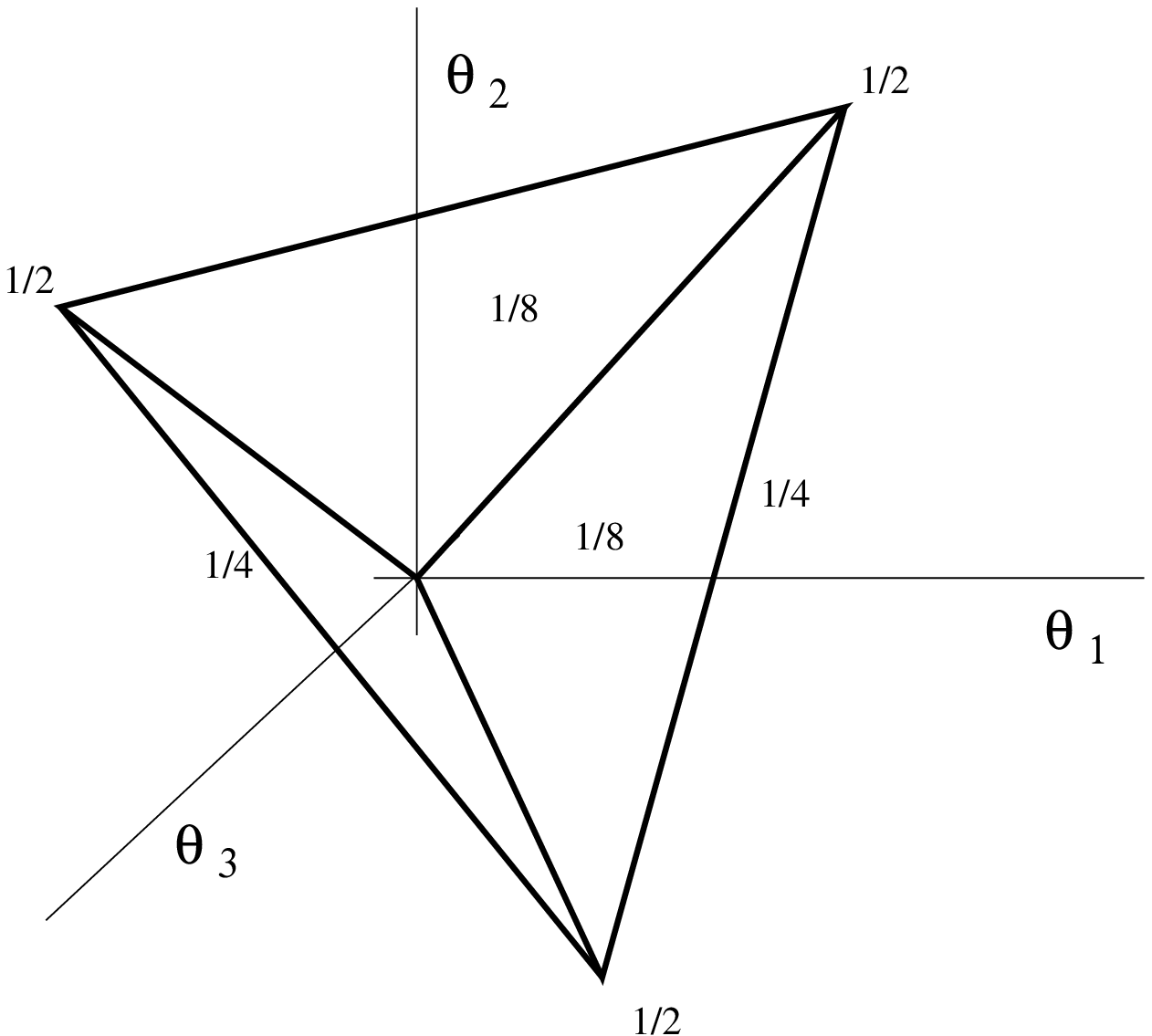, width=18.3em}{\label{T6}Angle parameter space 
for a system of two branes wrapping 3-cycles on $T^6$.}}

\noindent
Non-supersymmetric \linebreak[3] and non-tachyonic \\ 
(inside the tetrahedron), \\[2mm]
${\cal N}= 1$ supersymmetric \\
(faces of the tetrahedron), \\[2mm]
${\cal N}= 2$ supersymmetric \\
(edges of the tetrahedron), \\[2mm]
${\cal N}= 4$ supersymmetric \\
(vertices of the tetrahedron), \\ [2mm]   
Non-supersymmetric and tachyonic \\
(outside the tetrahedron). \\

Lower dimensional cases can be derived from this one by taking one
(four-dimensional torus) or two (two-dimensional torus) angles to
zero. In the two-dimensional case the system is always tachyonic,
signalling the instability of the system to the decay to a lower
volume brane. In the four-dimensional case, the system can be
supersymmetric or unstable, without the possibility of getting a
configuration of two branes that can be volume minimising.

\subsection{Critical points}

In the 2-dimensional case, the minima of the potential (\ref{pot}) can
be studied directly. In the 4-dimensional case, most of the
configurations can also be analysed directly. The most interesting
case is the 6-dimensional torus. As we have already mentioned, the
stable points of the NS-NS potential coincide with the final points of
the flow of the attractor equations considered in
Refs.~\cite{blackholes,Denef,Moore}. We will follow the analysis of
these equations done by Moore in Ref.~\cite{Moore}.

In particular, in a 6-dimensional space, Moore shows that if
$|Z_{\Gamma}(z)|$ has a stationary point in $z_*(\Gamma) \in {\cal M}$
with $|Z_{\Gamma}(z_*)| \neq 0$ then the 3-form dual to the cycle can
be decomposed as $\Gamma = \Gamma^{3,0} + \Gamma^{0,3}$. This
stationary point, if it is in the interior of the complex structure
moduli space, it must be a local minimum. Then, at the critical point,
$\Gamma^{3,0}$ should be proportional to the $\Omega$ form, up to a
phase, $\Gamma^{3,0} = -i \bar{C}\Omega$, where $C$ is a complex
number. Since $\Gamma \in H^3(X,\Z)$, then $\Gamma^{0,3} = i C
\bar{\Omega}$. The splitting condition above thus translates into
\begin{equation}\label{critical}
2\,\Im(\bar{C}\Omega) = \Gamma \,.
\end{equation}
Choosing a symplectic basis for $H^3(X,\Z)$, with internal product
$(\alpha_I, \beta^J) = \int \alpha_I \wedge \beta^J = \delta_I^{\ J}$,
we can write the cycle $\Gamma = p^I \alpha_I - q_J \beta^J$, where 
the coefficients $p_I, q_J$ are integers. The form can be written
as $\Omega = X^I \alpha_I + F_J \beta^J$, and the splitting condition 
becomes
\begin{eqnarray}
\bar{C} X^I - C\bar{X}^I & = & i p^I \nonumber \\
\bar{C} F^I - C\bar{F}^I & = & i q^I \,,
\end{eqnarray}
where $X^I$ and $F^I$ are the periods along the $\alpha_I$ and
$\beta^I$ cycles, respectively, $X^I = \int_{\alpha_I}\Omega = \int
\Omega \wedge \beta^I$. These are $b_3$ equations for $b_3$ real
variables (where $b_3$ is the dimension number of $H^3(X,\R)$), so we
can expect the solutions to be isolated points in the complex
structure moduli space.

\section{The two-dimensional torus}

In the case of two-dimensional tori, the special Lagrangian
submanifolds are straight lines in the covering space of the torus.
There is one for each homology class $\{n [a] + m [b]\}$. The moduli
of these curves is ${\cal M}_\Sigma = \R$ and correspond to
translations in the transverse directions to the branes. They can be
complexified if Wilson lines are taken into account, see for
instance Ref.~\cite{Denef}.

In this case, the holomorphic 1-form is $\Omega_0 = dz$, where $dz =
dx + \tau dy$, and $\Im\,\tau > 0$. The metric on the torus is
$ds^2 = dz d\bar{z}$ and the K\"ahler form is $\omega = dz \wedge
d\bar{z}$. The volume of the torus then becomes
\begin{equation}
{\rm Vol} = i \int_{T^2} \Omega_0 \wedge \bar\Omega_0 = \Im\,\tau\,.
\end{equation}
The K\"ahler potential for the complex structures is then
${\cal K} = -\ln({\rm Vol}) = -\ln(\Im\,\tau)$.
The K\"ahler metric in the half plane of complex structures becomes
\begin{equation}
ds^2 = \frac{d\tau d\bar{\tau}}{(\Im\,\tau)^2}\,,
\end{equation}
and the normalised 1-form:
\begin{equation}
\Omega \equiv e^{{\cal K}/2} \Omega_0 = {dx + \tau dy\over
\sqrt{\Im\,\tau}}  \,.
\end{equation}
The periods of the cycles where the D-branes are wrapped become
\begin{equation}
Z_{\Gamma} = \int_{\Gamma} \Omega\,,
\end{equation}
which has the interpretation of the volume of the cycle relative to the
square root of the volume of the whole torus.

The potential obtained from the NS-NS tadpoles is related to the
periods of the brane wrapping the cycle $\Gamma$ by Eq.~(\ref{vola}).
In this case we have
\begin{equation}
V(\phi,\tau) = e^{- \phi} \frac{|n + \tau m|}{\sqrt{\Im\,\tau}} \,.
\end{equation}

In the two-dimensional torus we can distinguish two cases:

\begin{itemize}
  
\item if $m=0$, i.e. a brane only wrapping the $[a]$ cycle, the minimum is
  at $\Im\,\tau \rightarrow \infty$, see Fig.~\ref{T210}.
  
  \smallskip \DOUBLEFIGURE[b]{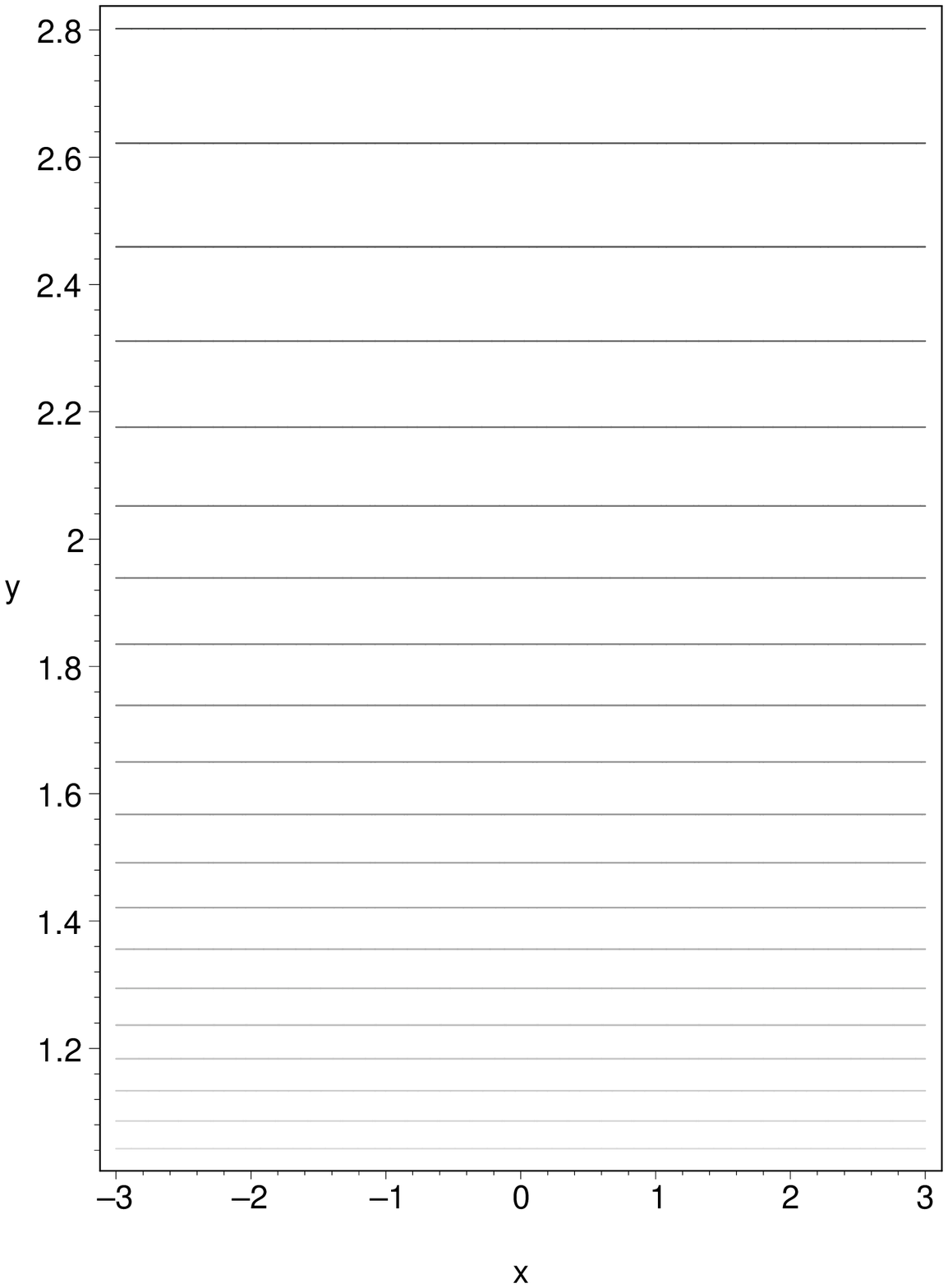, width=.4\textwidth} {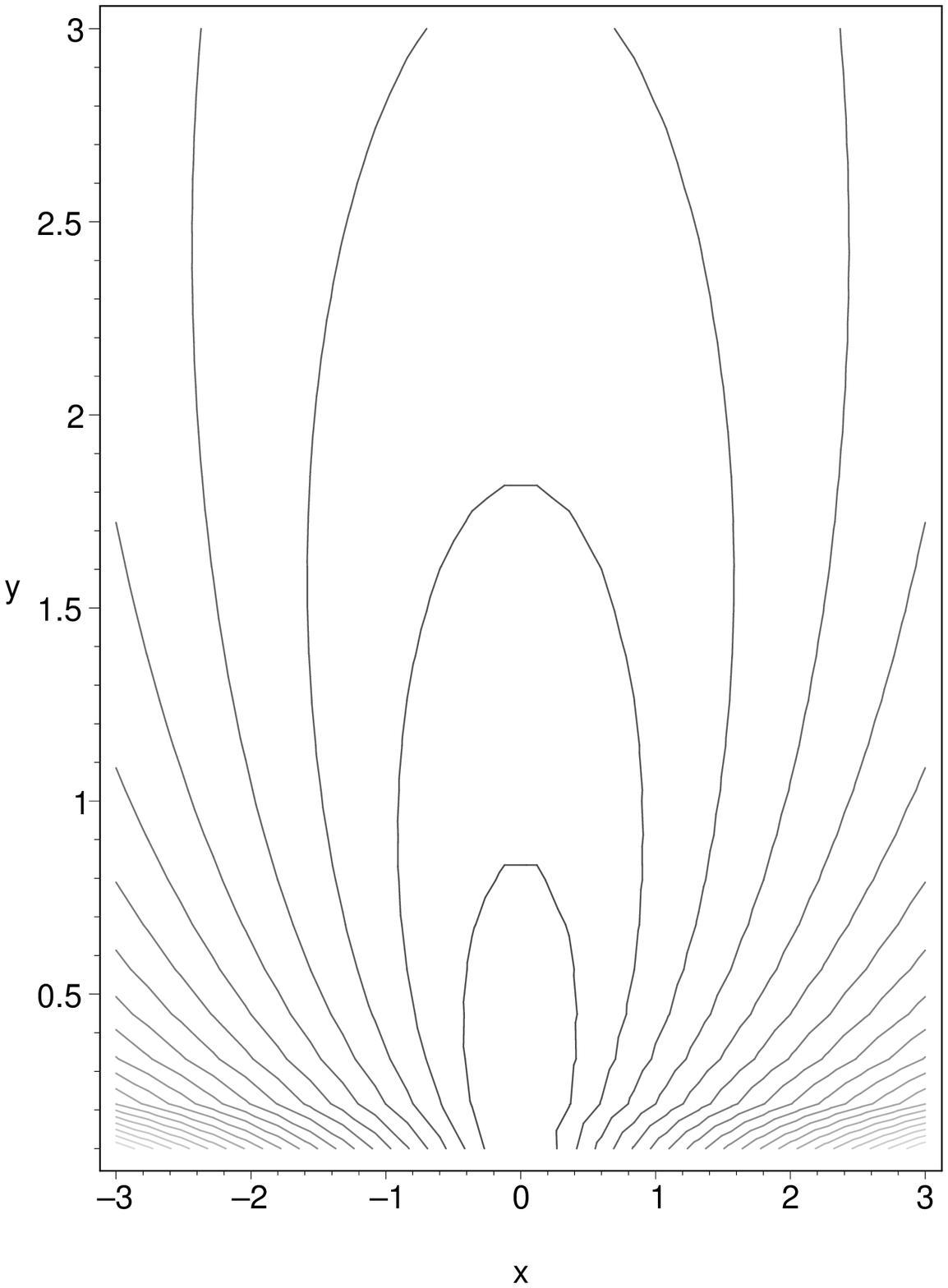,
    width=.4\textwidth}{\label{T210}Contour plot for the potential
    generated by a brane wrapping the $(1,0)$ cycle in a two
    dimensional torus.}{\label{T201}Contour plot for the potential
    generated by a brane wrapping the $(0,1)$ cycle in a two
    dimensional torus.}
  
\item if $m \neq 0$ the minimum is at $\tau \rightarrow -n/m$, a real
  number, see Fig.~\ref{T201}.

\end{itemize}

\smallskip \DOUBLEFIGURE[b]{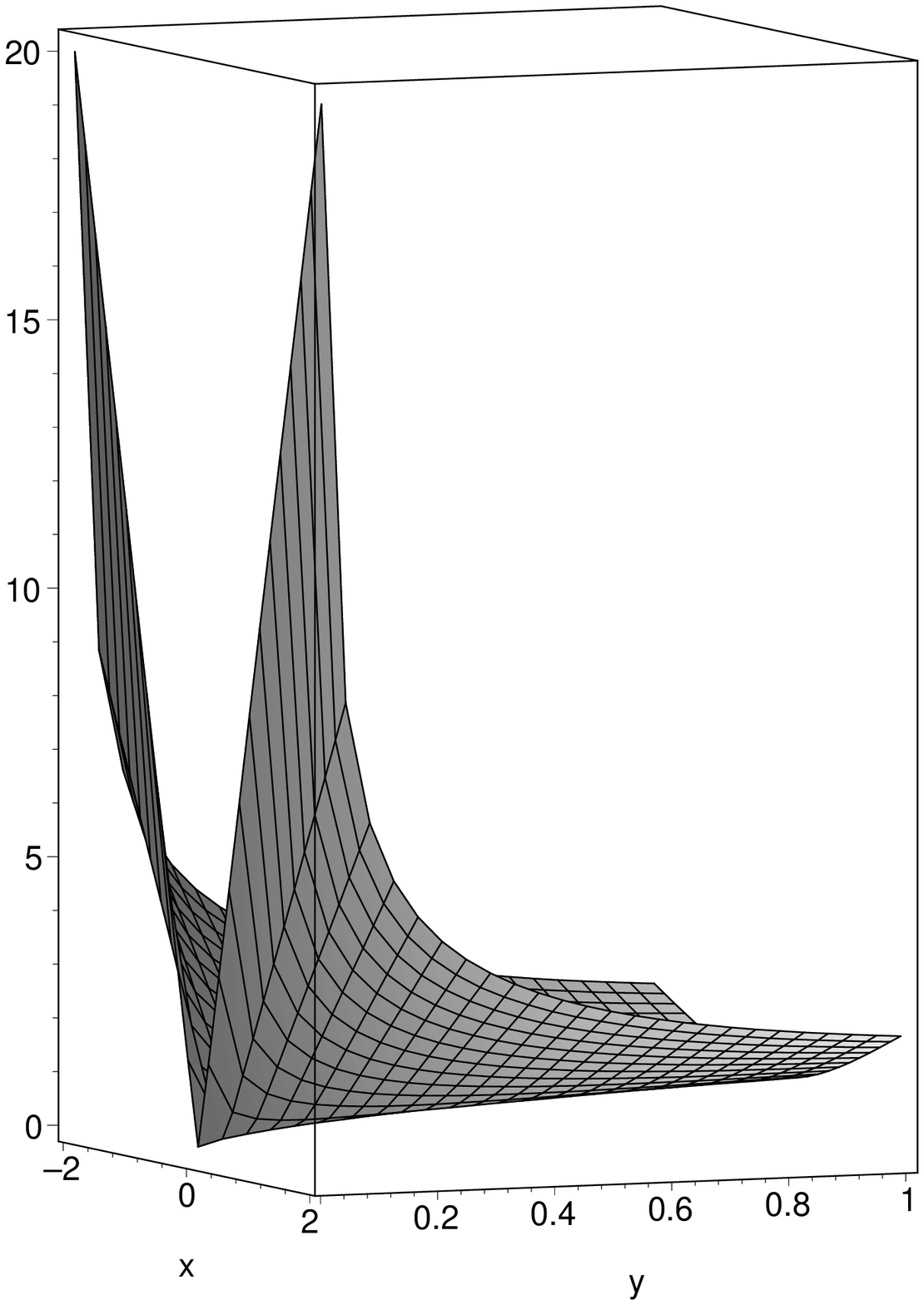, width=.4\textwidth}
{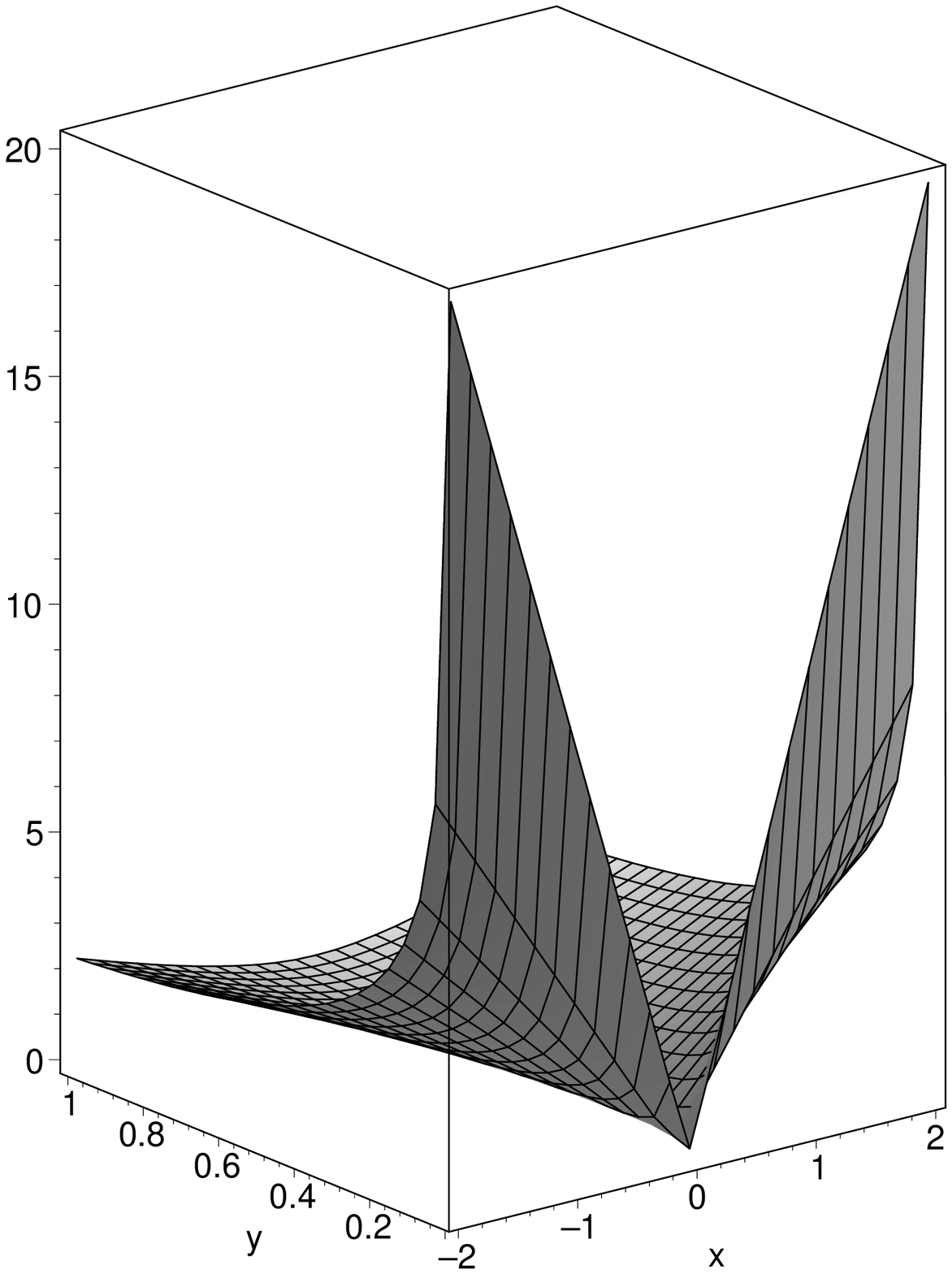, width=.4\textwidth}{\label{T210detalle1}Plot of the
  potential generated by a brane wrapping the $(1,0)$ cycle in a two
  dimensional torus.}{\label{T210detalle2}Plot of the potential
  generated by a brane wrapping the $(1,0)$ cycle in a two dimensional
  torus.}

In both cases the system is driven by this potential to the boundary
of the complex structure moduli space, where the volume of the cycle where
the brane is wrapped goes to zero. The brane is stable against decays
into other type of branes.

The Lagrangian for the complex structure moduli is of the form
\begin{equation}
{\cal L} = e^{-2 \phi} \frac{\partial_{\mu}\tau 
\partial^{\mu}\bar{\tau}}{(\Im\,\tau)^2} - e^{- \phi} 
\frac{|n + \tau m|}{\sqrt{\Im\,\tau}}\,.
\end{equation}
By performing a T-duality along the $(1,0)$ direction one can
understand this flow as the one responsible for the contraction of the
manifold to a point when the D-brane wraps the whole manifold, or its
expansion, when T-duality takes the brane to a lower dimensional one,
as already mentioned in the introduction.

\section{The four-dimensional torus}

In this case, the holomorphic 2-form of the 4-dimensional torus is
$\Omega_0 = dz_1 \wedge dz_2$, where $dz_i = dx_i + \tau_{ij} dy_i$
and $\tau_{ij}$ is a 2x2 complex matrix that characterises the complex
structure of the torus.  The metric on the torus is $ds^2 = \sum_i
dz_i d\bar{z}_i$, and the K\"ahler form, $\omega = \sum_i dz_i \wedge
d\bar{z}_i$. The volume of the torus becomes
\begin{equation}
{\rm Vol} =  \int_{T^4} \Omega_0 \wedge \bar\Omega_0 =  
{\rm det}\,\tau + {\rm det}\,\bar{\tau} - \tau_{11} \bar{\tau}_{22} 
- \tau_{22} \bar{\tau}_{11} + \tau_{12} \bar{\tau}_{21} + 
\tau_{21} \bar{\tau}_{12}\,.
\end{equation}
The K\"ahler potential for the complex structures is as usual, ${\cal K}
= -\ln({\rm Vol})$. The K\"ahler metric in the plane of complex
structures, $g_{ij} = \partial_i \partial_j {\cal K}$.  The normalised
2-form becomes $\Omega \equiv e^{{\cal K}/2} \Omega_0$. Now we have
the 2-cycles dual to the forms $dx^1 \wedge dx^2$, $dx^i \wedge dy^j$,
$dy^1 \wedge dy^2$ that form a basis of $H^2(X,\R)$. Let us denote the
wrapping numbers along these cycles by $q_0$, $q_{ij}$, $\tilde{q}_0$.
The periods of the cycles where the branes are wrapped are given by
\begin{equation}
Z_{\Gamma} = \int_{\Gamma} \Omega = {q_0 + q_{ij} \tau_{ij} + 
\tilde{q}_0\,{\rm det}\,\tau \over\sqrt{\rm Vol}}\,,
\end{equation}
which has the interpretation of the volume of the cycle relative to
the square root of the volume of the whole manifold. The potential
from the NS-NS tadpoles are related to the periods by $V(\phi,\tau) =
e^{- \phi} |Z_{\Gamma}|$. Some interesting cases are:

\

\noindent
a) If the metric factories into two 2-dimensional tori, i.e. $\tau_{ij}
= \delta_{ij} \tau_i$, then the volume is 
\begin{equation}
{\rm Vol} = \prod_i \Im\,\tau_i\,,
\end{equation}
and the potential takes a very simple form,
\begin{equation}
V(\phi,\tau) = e^{- \phi} {|q_0 + \tau_1 q_{11} + \tau_2 q_{22} 
+ \tau_1 \tau_2 \tilde{q}_0|\over\prod_i \sqrt{\Im\,\tau_i}}\,.
\end{equation}
Note that in this case we are in a point in the complex structure
moduli space where some cycles have zero volume, those with coordinates
$q_{12}$ and $q_{21}$.  Now let us consider the following subcases:

\

\noindent
a.1) If the cycle is also factorizable into two 1-cycles, each one
wrapping a two-dimensional torus, then we can denote these 1-cycles by
$(r_1,s_1)$ and $(r_2,s_2)$. The potential is now
\begin{equation}
V(\phi,\tau) = e^{- \phi} \prod_i \frac{|r_i + s_i \tau_i|}
{\sqrt{\Im\,\tau_i}}\,.
\end{equation}
The problem of analysing this potential reduces to that of the
two-dimensional torus. The system is then driven to the boundaries of
the complex structure moduli where the 1-cycles collapse.

\

\noindent
a.2) We do not consider the cycle factorizable but we keep the
same complex structure in both two-dimensional tori, i.e. $\tau_1 =
\tau_2 = \tau$ . Let us define $q \equiv q_{11} + q_{22}$. Then the
potential becomes
\begin{equation}
V(\phi,\tau) = e^{- \phi} \frac{|q_0 + \tau q + 
\tau^2 \tilde{q}_{0}|}{\Im\,\tau} \,.
\end{equation}
The behaviour of this potential is determined by the sign of the
discriminant, $\ \Delta = q^2 - 4 q_0 \tilde{q}_{0}$, 
of the polynomial:
\begin{equation}
p(\tau) = q_0 + \tau q + \tau^2 \tilde{q}_{0}  
\end{equation}
The different cases are:

\

a.2.i) If $\Delta > 0$, then the two roots are real and are at the
boundary. The minimum is in a line joining the two roots. The value of
the minimum of the potential is different from zero, $V_0(\phi) = e^{-
  \phi} \Delta / \tilde{q}^2_{0}$. See Figs.~\ref{T4-101} and
\ref{T4-1013D}. Note that the factorizable cycle cases are of this
type.

\smallskip \DOUBLEFIGURE[b]{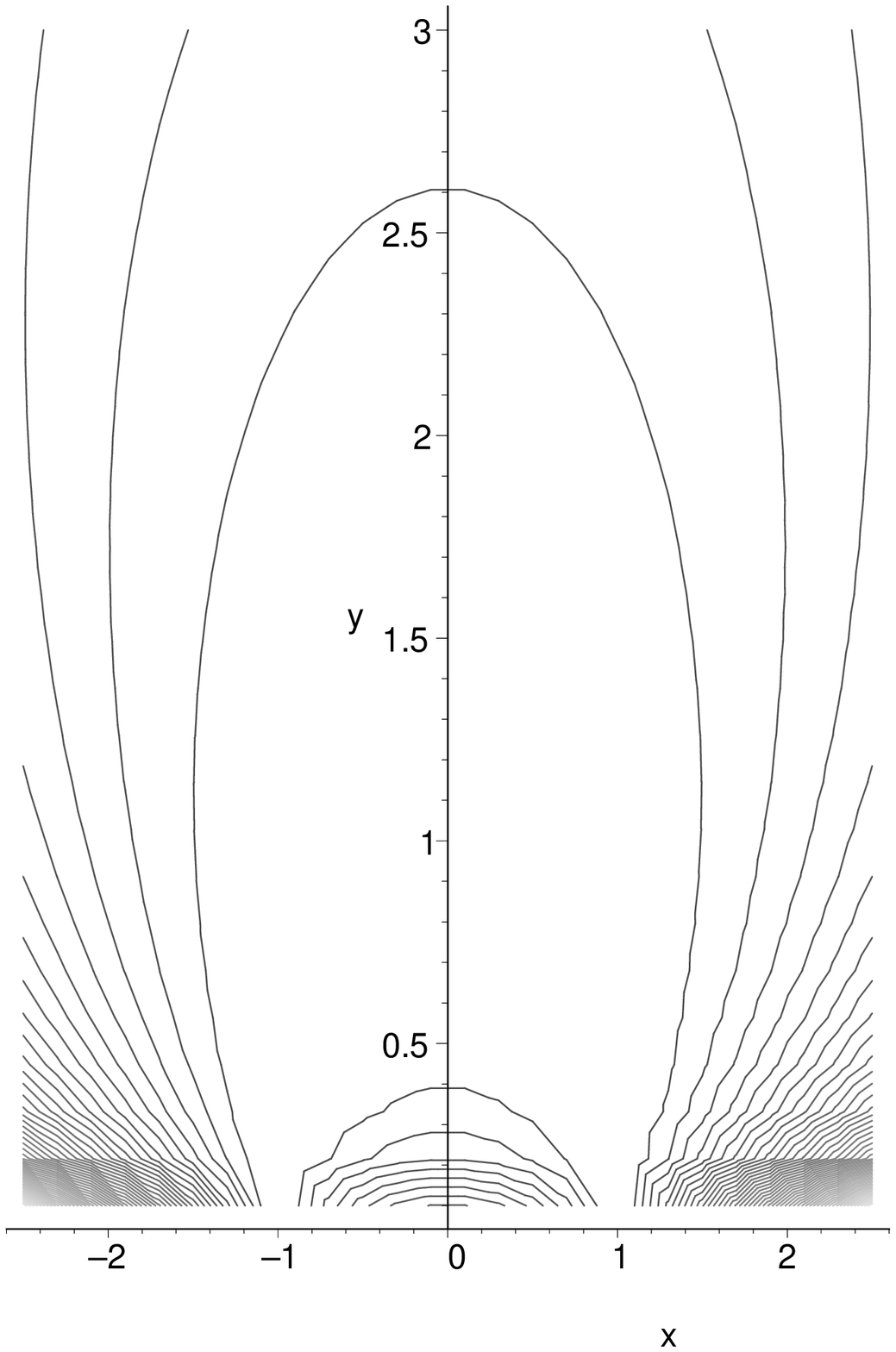, width=.4\textwidth}
{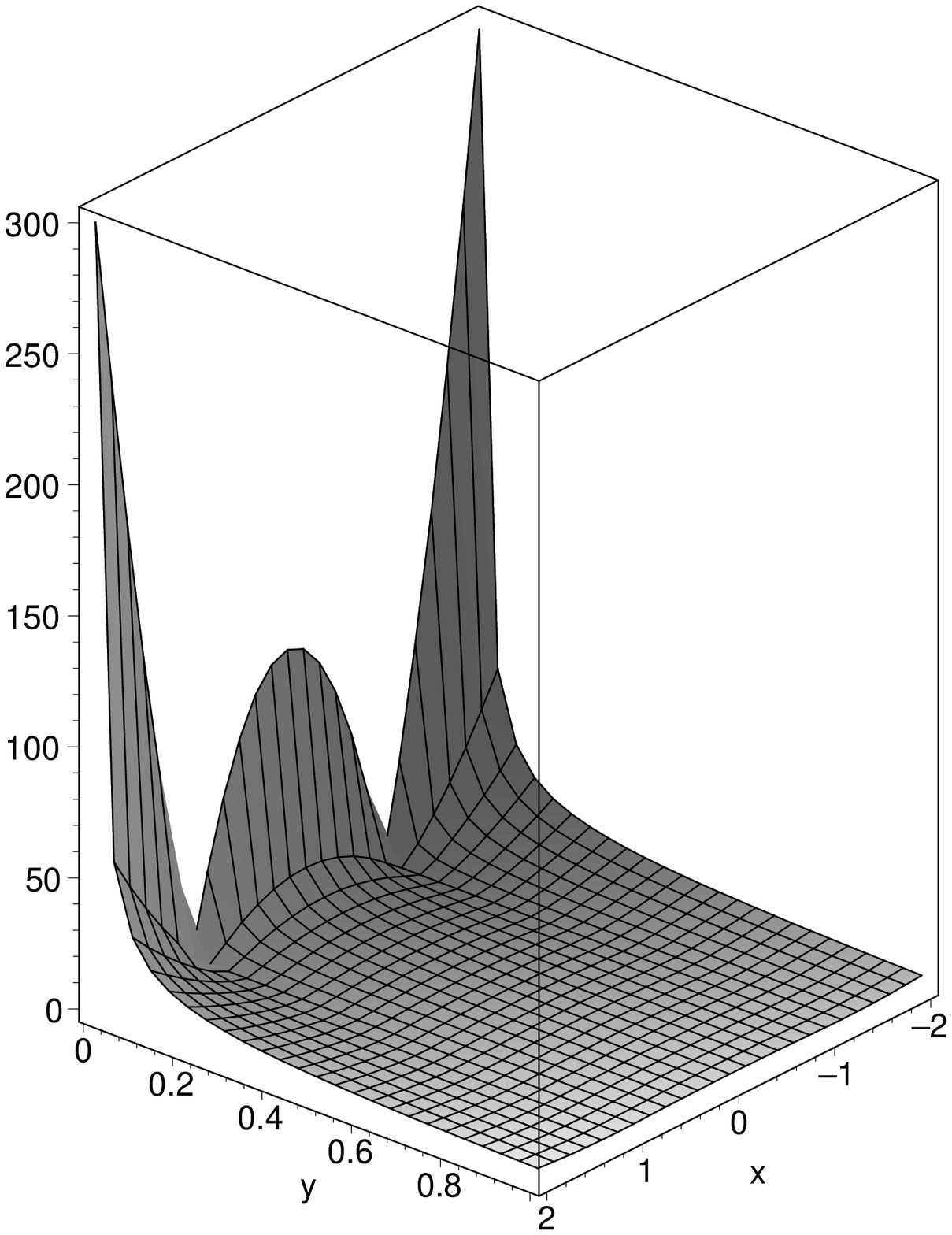, width=.4\textwidth}{\label{T4-101}Contour plot of the
  potential generated by a brane wrapping the $q_0 = -1$, $q = 0$,
  $\tilde{q}_0 =1$ cycle in a four dimensional factorizable torus with
  the same complex structure in the two 2-tori.}{\label{T4-1013D}Three
  dimensional representation of the potential generated by a brane
  wrapping the $q_0 = -1$, $q = 0$, $\tilde{q}_0 =1$ cycle in a four
  dimensional factorizable torus with the same complex structure in
  the two 2-tori.}

\

a.2.ii) If $\Delta = 0$, then the two roots are real and coincide. The
minimum is at the root, in the boundary. The value of the minimum of
the potential is at zero. See Figs.~\ref{T4001} and \ref{T40013D}.

\smallskip \DOUBLEFIGURE[b]{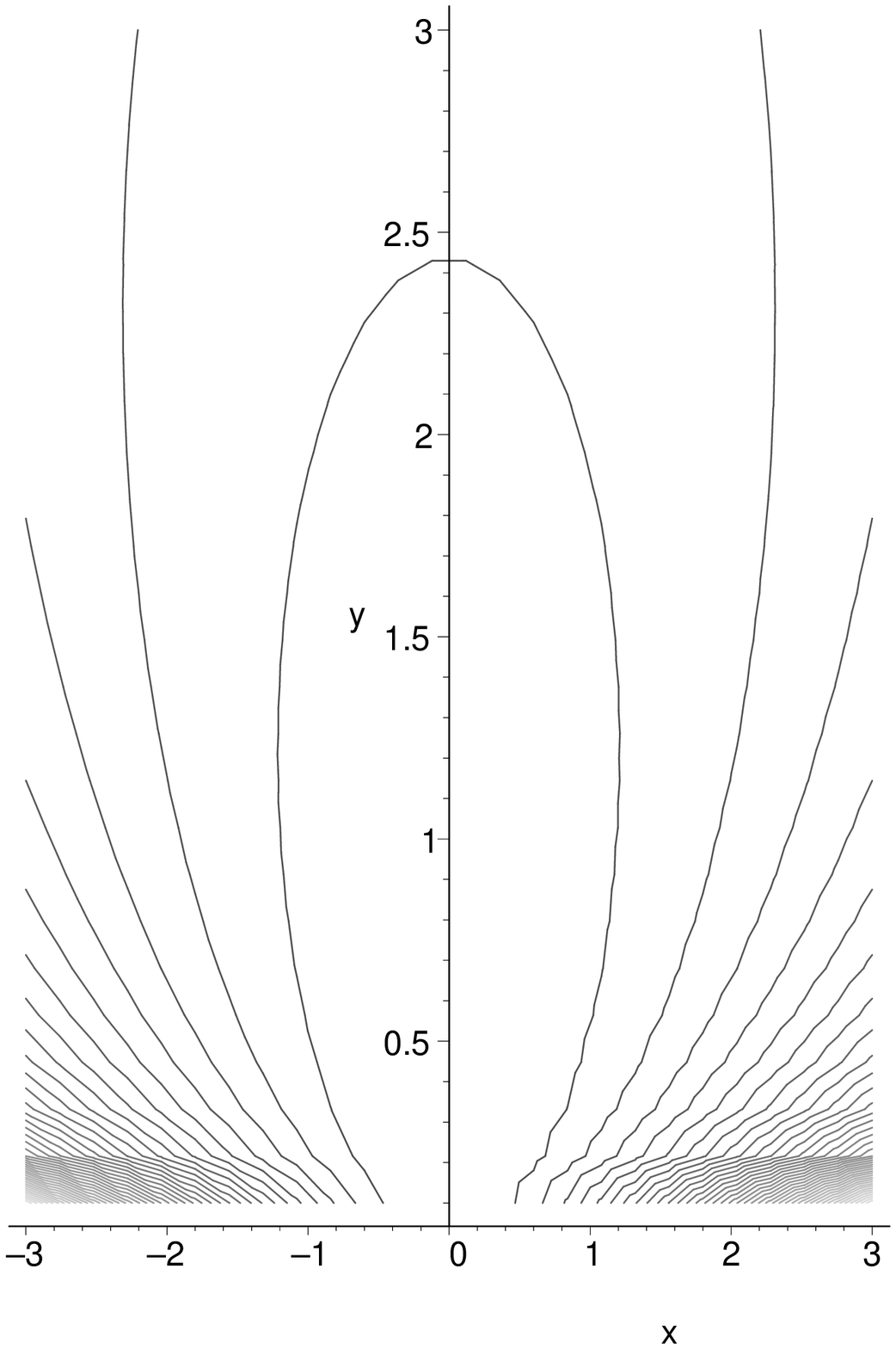, width=.4\textwidth}
{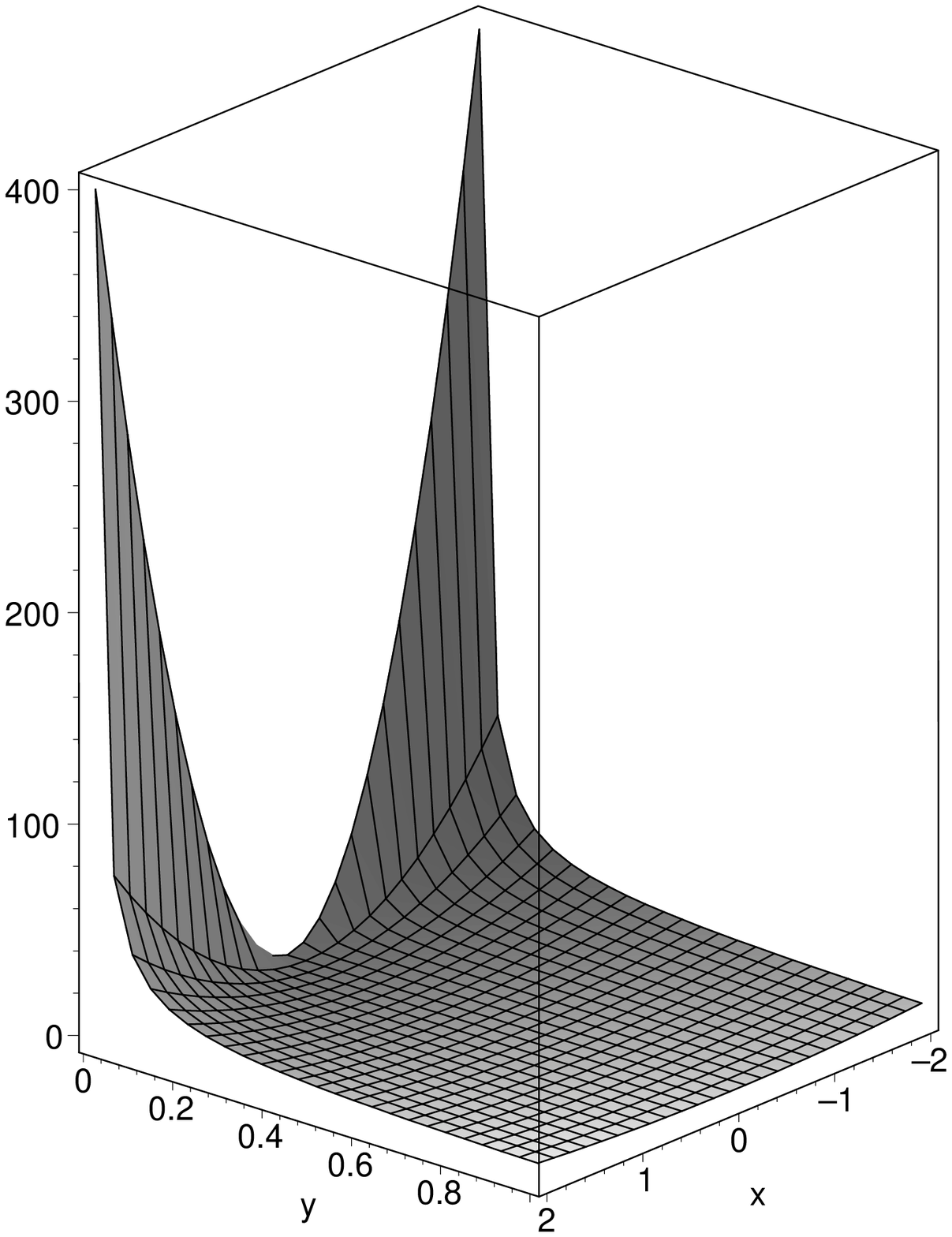, width=.4\textwidth}{\label{T4001}Contour plot of the
  potential generated by a brane wrapping the $q_0 = 0$, $q = 0$,
  $\tilde{q}_0 = 1$ cycle in a four dimensional factorizable torus
  with the same complex structure in the two 2-tori.}
{\label{T40013D}Three dimensional representation of the
  potential generated by a brane wrapping the $q_0 = 0$, $q = 0$,
  $\tilde{q}_0 = 1$ cycle in a four dimensional factorizable torus
  with the same complex structure in the two 2-tori.}

\

a.2.iii) If $\Delta < 0$, then the two roots are complex conjugates.
The minimum is at the root, in the interior of the moduli space of
complex structures. The value of the minimum of the potential is at
zero. Following the analysis of Moore \cite{Moore}, it seems that
there is no BPS state at this point. We will see in some specific
examples that this is indeed the case. When $\Delta < 0$ the system
will cross a line of marginal stability and the brane is expected to
decay into another system. Note that this will never be the case when
the cycle is factorizable. See Figs.~\ref{T4101} and \ref{T41013D}.
We will analyse examples of line-crossing in the more interesting 
case of 6-dimensions.

\smallskip \DOUBLEFIGURE[b]{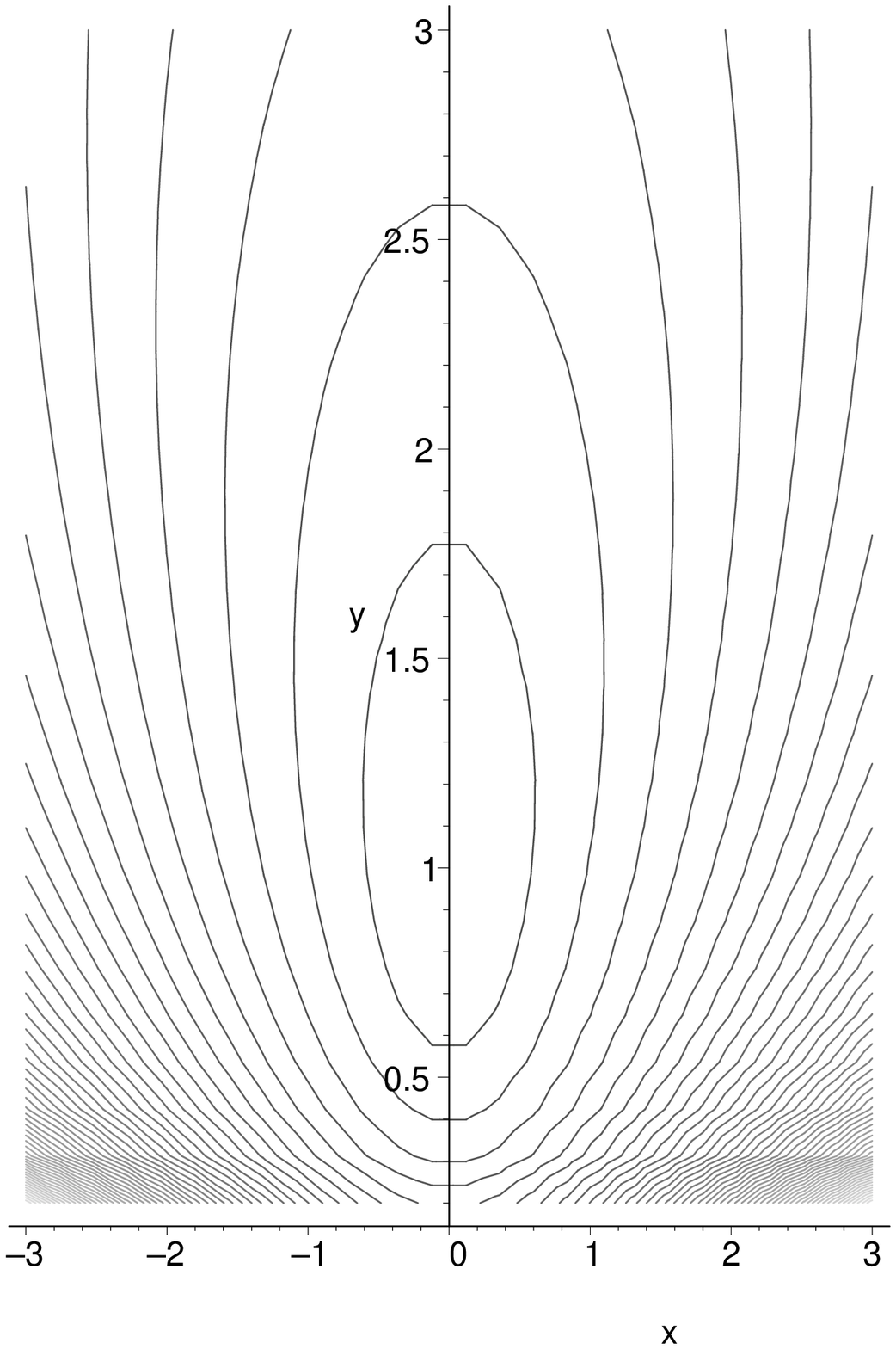, width=.4\textwidth}
{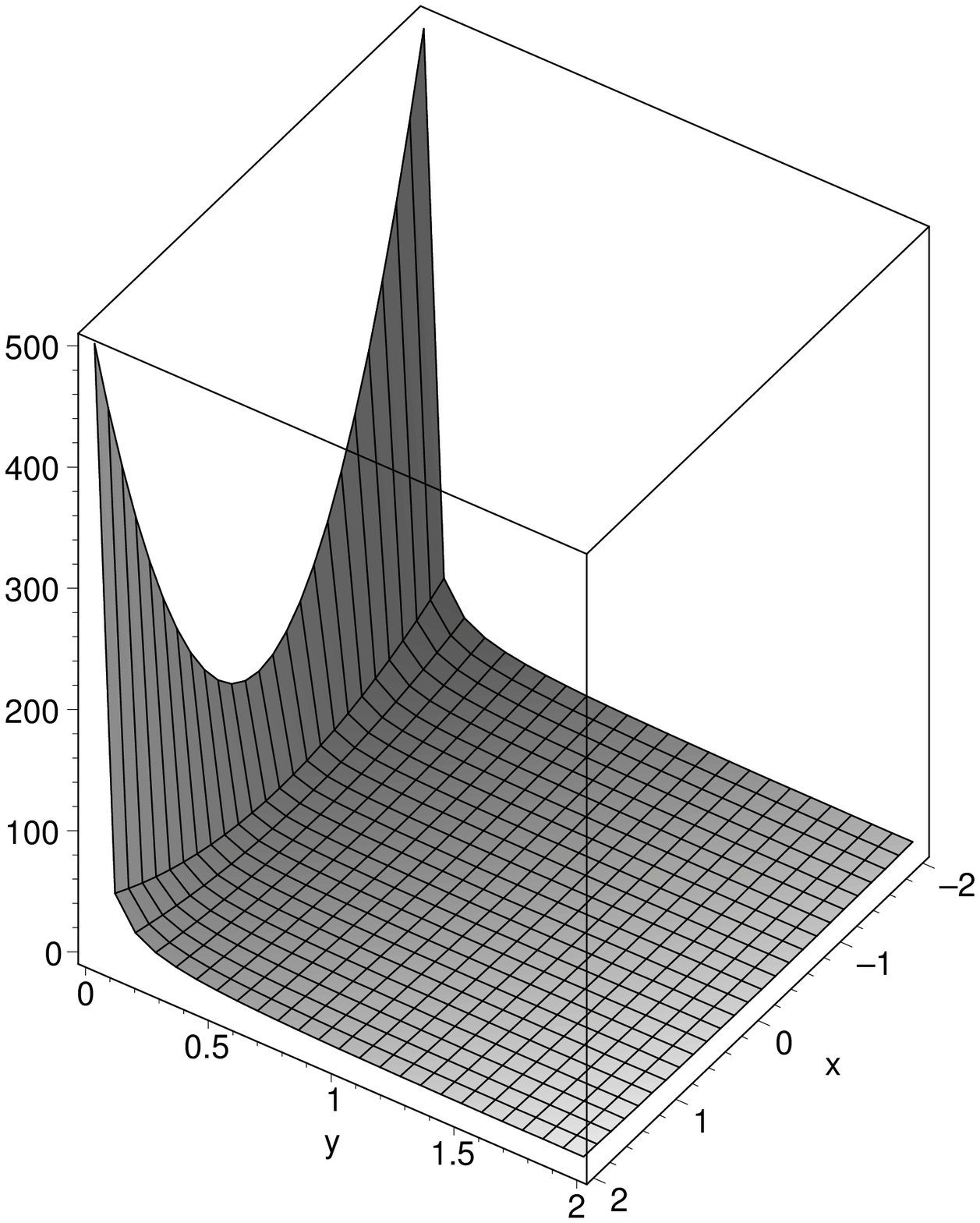, width=.4\textwidth}{\label{T4101}Contour plot of the
  potential generated by a brane wrapping the $q_0 = 1$, $q = 0$,
  $\tilde{q}_0 = 1$ cycle in a four dimensional factorizable torus
  with the same complex structure in the two 2-tori.}
  {\label{T41013D}Three dimensional representation of the
  potential generated by a brane wrapping the $q_0 = 1$, $q = 0$,
  $\tilde{q}_0 = 1$ cycle in a four dimensional factorizable torus
  with the same complex structure in the two 2-tori.}

\

\noindent
b) The general case in which the complex structure part of the metric
does factorise will not be analysed here. Naive extrapolation from the
6-dimendional analysis ($T^6 = T^4\times T^2$) indicates that the
system is driven to the boundary ($\Delta=0$). This was expected,
since there is a 2-dimensional torus that has always this behaviour.

\section{The six-dimensional torus}

In this case, the holomorphic 3-form is $\Omega_0 = dz_1 \wedge dz_2
\wedge dz_3$, where $dz_i = dx_i + \tau_{ij} dy_i$.  The metric on the
6-torus is defined by $ds^2 = \sum_i dz_i d\bar{z}_i$ and the K\"ahler
form becomes $\omega = \sum_i dz_i \wedge d\bar{z}_i$. The volume of
the torus is
\begin{equation}
{\rm Vol} =  i \int_{T^6} \Omega_0 \wedge \bar\Omega_0 =  
i \left[{\rm det}\,\tau - {\rm det}\,\bar{\tau} + 
{\rm tr}(\tau\,{\rm Cof}\,\bar\tau) -  
{\rm tr}(\bar{\tau}\,{\rm Cof}\,\tau)\right]\,,
\end{equation}
where the cofactor of a matrix is ${\rm Cof}A = {\rm det}A\,
(A^{-1})^T$.  The K\"ahler potential for the complex structures is 
${\cal K} = -\ln({\rm Vol})$. The K\"ahler metric in the plane of complex
structures, $g_{ij} = \partial_i \partial_j {\cal K}$. The normalised
3-form: $\Omega \equiv e^{{\cal K}/2} \Omega_0$. Now we have the
3-cycles dual to the following forms, which form a basis of $H^3(T^6,\R)$,
\begin{equation}
\begin{array}{rl}
\alpha_0 & = dx^1 \wedge dx^2  \wedge dx^3\,, \nonumber \\
\alpha_{ij} & = \half\epsilon_{ilm}\, dx^l \wedge dx^m  
\wedge dy^j\,, \nonumber \\
\beta^{ij} & = \half\epsilon_{jlm}\, dx^i \wedge dy^l  
\wedge dy^m\,, \nonumber \\
\beta_0 & = - dy^1 \wedge dy^2  \wedge dy^3\,,
\end{array}
\end{equation}
which satisfy the relation:
\begin{equation}
\int_{T^6} \alpha_I \wedge \beta^J = \delta_I^{\ J}\,.
\end{equation}
The wrapping numbers along these cycles are $q_0$, $Q_{ij}$, $P^{ij}$,
$p^0$, respectively. And the periods of the cycles where the branes
are wrapped can be written as
\begin{equation}
Z_{\Gamma} = \int_{\Gamma} \Omega = {q_0 + Q_{ij}\,\tau^{ij} + 
P^{ij}\,{\rm Cof}\,\tau_{ij} - p^0\,{\rm det}\,\tau\over\sqrt{\rm Vol}}\,,
\end{equation}
which has the interpretation of the volume of the cycle relative to
the square root of the volume of the whole manifold.  The potential from
the NS-NS tadpoles are related to the periods by $V(\phi,\tau) = e^{-
\phi} |Z_{\Gamma}|$. Particular cases are:

\ 

\noindent
a) If the metric factories into three 2-dimensional tori, i.e. 
$\tau_{ij} = \delta_{ij} \tau_i$, then the volume is 
\begin{equation}
{\rm Vol} = \prod_i \Im\,\tau_i \,,
\end{equation}
and the potential takes a very simple form,
\begin{equation}
V(\phi,\tau) = e^{- \phi} {\left|q_0 + \sum_i Q_{ii}\,\tau^i + 
\half \sum_i P^{ii} \epsilon_{ijk} \tau^j\tau^k - 
p^o\tau^1\tau^2\tau^3\right|\over\prod_i \sqrt{\Im\,\tau_i}} 
\end{equation}
Note that in this case we are at a point in the complex structure
moduli space where some cycles have zero volume, those with coordinates
$Q_{ij}$ and $P^{ij}$, with $i \neq j$.  Now let us consider the following
subcases:

\noindent
a.1) If the cycle is also factorizable into two 1-cycles, each one
wrapping a two dimensional torus. Let us denote these 1-cycles by
$(r_1,s_1)(r_2,s_2)(r_3,s_3)$. The potential is now:
\begin{equation}
V(\phi,\tau) = e^{- \phi} \prod_i \frac{|r_i + s_i \tau_i|}
{\sqrt{\Im\,\tau_i}} 
\end{equation}

The problem of analysing this potential reduces to the two dimensional
torus problem. The system is then driven to the boundaries of the
complex structure moduli where the 1-cycles collapse.

\

\noindent
a.2) We do not consider a factorizable cycle, but we keep the same
complex structure in all two-dimensional tori, i.e. $\tau_i = \tau$ .
Let us define $3 q \equiv \sum_i Q_{ii}$ and $3 p \equiv \sum_i P^{ii}$. Then
the potential becomes
\begin{equation}
V(\phi,\tau) = e^{- \phi} \frac{|q_0 + 3 q \tau  + 3 p \tau^2 - 
p^0 \tau^3|}{(\Im\,\tau)^{3/2}} 
\end{equation}
The behaviour of this potential is determined by the sign of the
discriminant, $\Delta = 12 p^2 q^2 - (3pq + p^0 q_0)^2 + 4 (p^0 q^3 -
q_0 p^3)$, of the polynomial:
\begin{equation}
p(\tau) = q_0 + 3 q \tau  + 3 p \tau^2 - p^0 \tau^3\,.
\end{equation}
The discriminant gives the number and the type of solutions to
$p(\tau)$. As in the four dimensional case, there are 3 subcases:

\

a.2.i) if $\Delta > 0$, there are three real roots, all different. The
minimum is in the interior of the complex structure moduli space. The
minimum of the potential is not vanishing. Following the
interpretation of Moore \cite{Moore}, this means that the
corresponding BPS state must exist. See Figs. \ref{T60-101} and
\ref{T60-1013D}. Note that this possibility can be achieved with a
factorizable cycle. The analysis seems to be in contradiction with the
case a.1). But now we are doing a partial analysis by considering all
the complex structures equivalent.

However one can get this kind of configurations by taking three
factorizables cycles. For example, take $(-1,0)(1,0)(1,0)$,
$(0,-1)(0,1)(0,1)$ and $(1,1)(1,1)(1,1)$. We will see this example in
detail in the last section.

\smallskip \DOUBLEFIGURE[b]{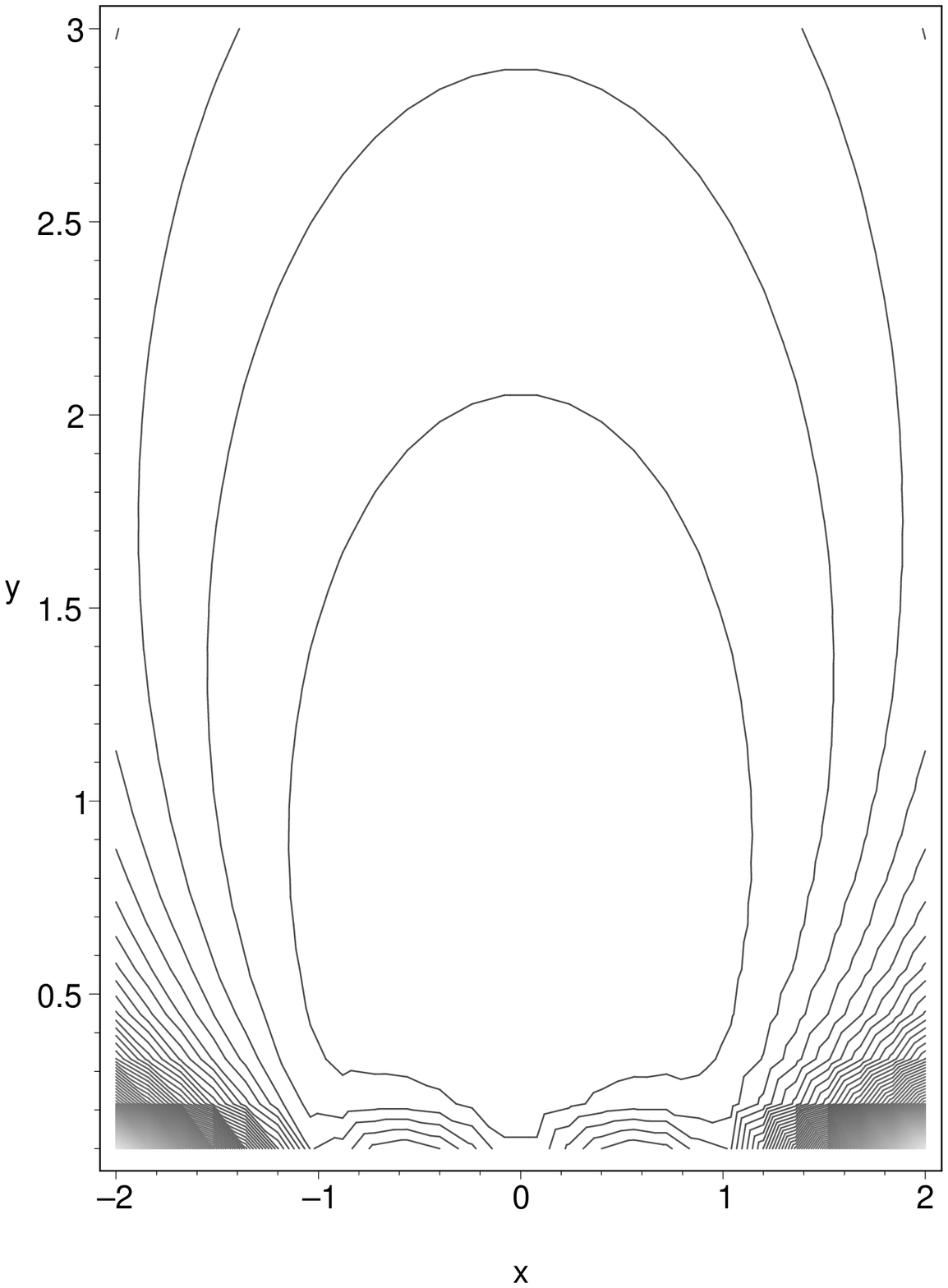, width=.4\textwidth}
{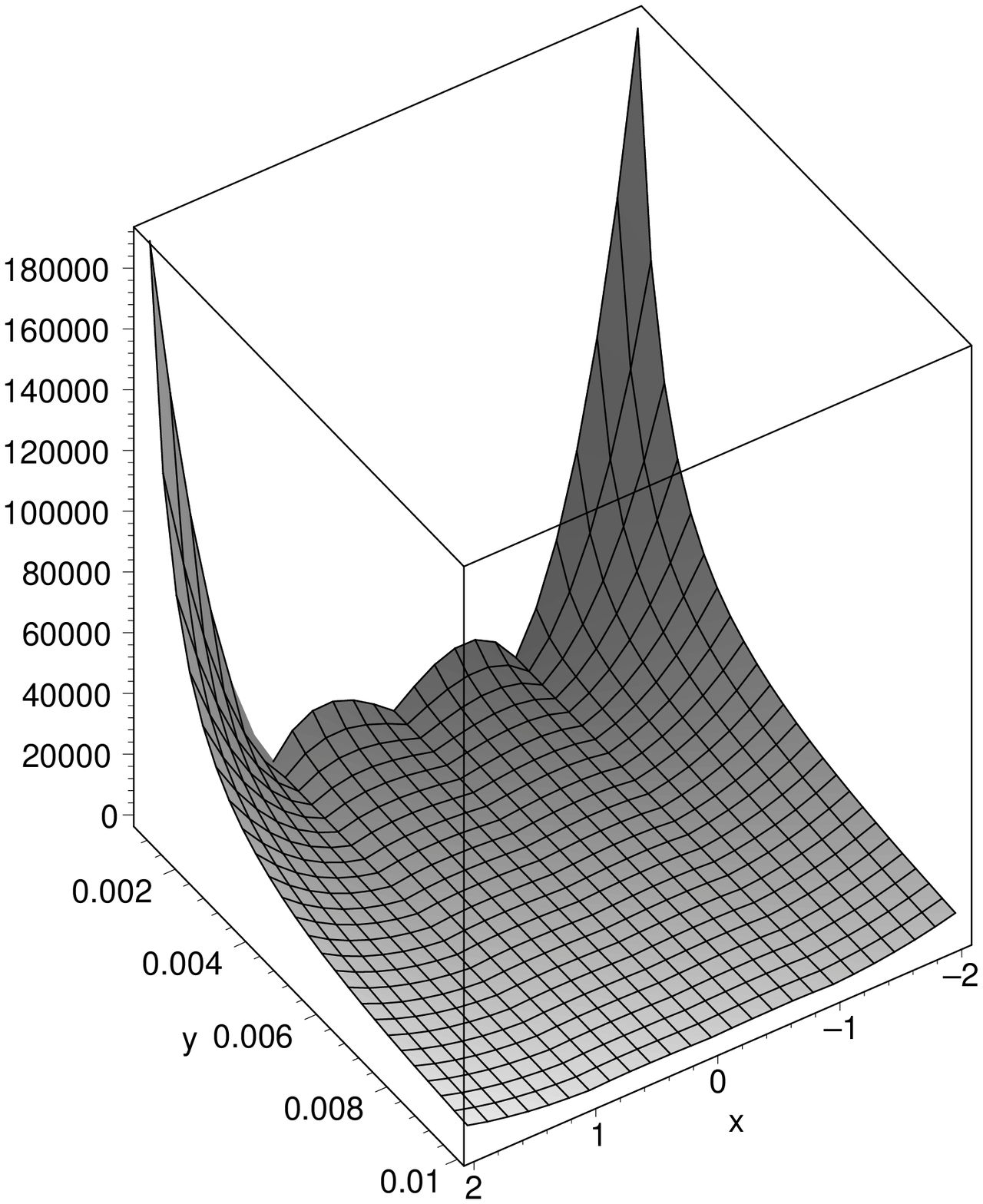, width=.4\textwidth}{\label{T60-101}Contour plot of the
  potential generated by a brane wrapping the $q_0 = 0$, $q = 1/3$, $p
  = 0$, $p^0 = -1$ cycle in a six dimensional factorizable torus with
  the same complex structure in the three 2-tori.}
  {\label{T60-1013D}Three dimensional representation of the
  potential generated by a brane wrapping the $q_0 = 0$, $q = 1/3$, $p
  = 0$, $p^0 = -1$ cycle in a six dimensional factorizable torus with
  the same complex structure in the three 2-tori.}

\

a.2.ii) if $\Delta = 0$, there are three real roots, but two of them
are equal.  The minimum is at the boundary. The potential goes to zero
at that point in the boundary. See Figs.~\ref{T60011} and
\ref{T600113D}. Notice that this possibility can be achieved with a
factorizable cycle.

\smallskip \DOUBLEFIGURE[b]{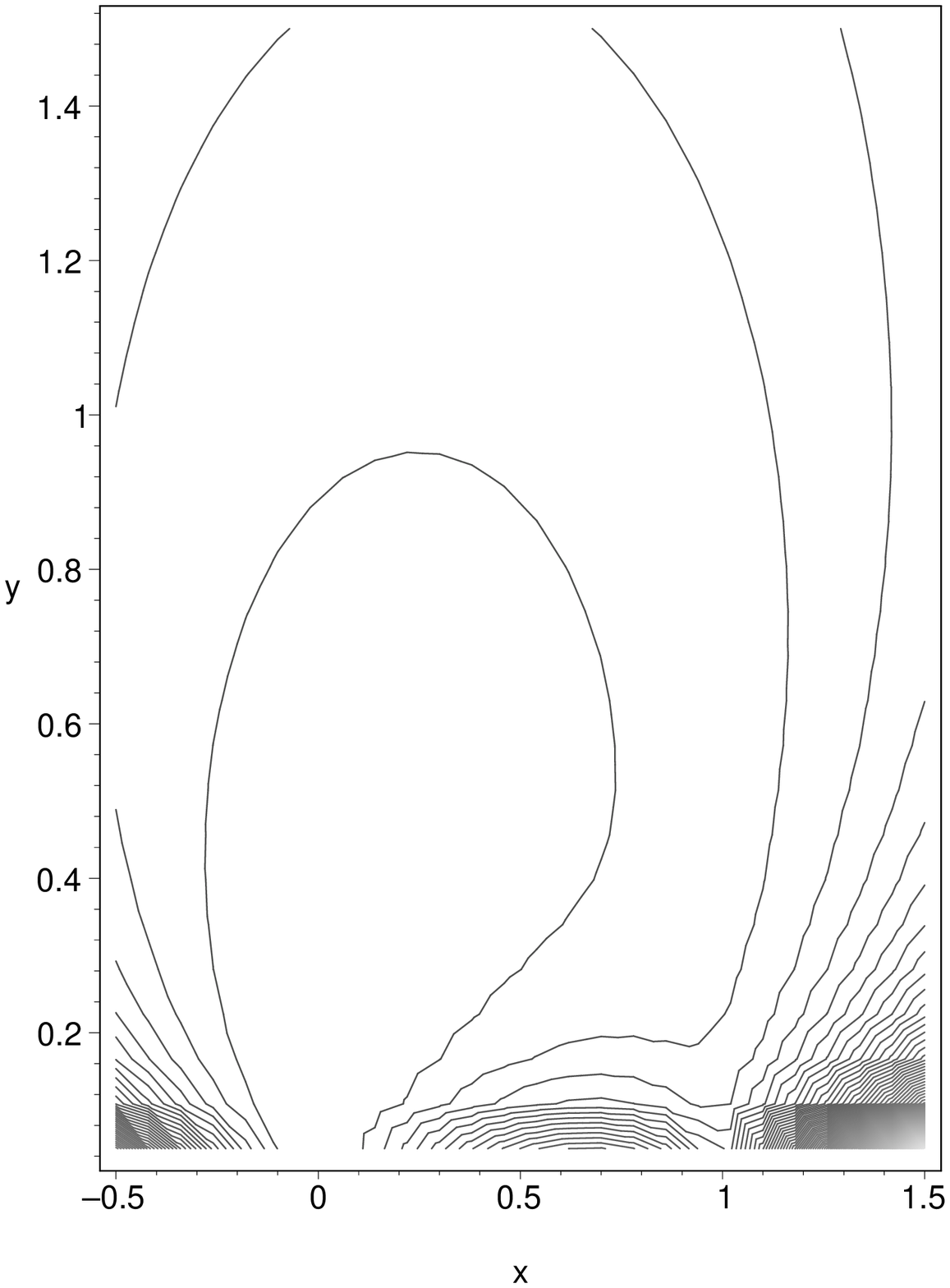, width=.4\textwidth}
{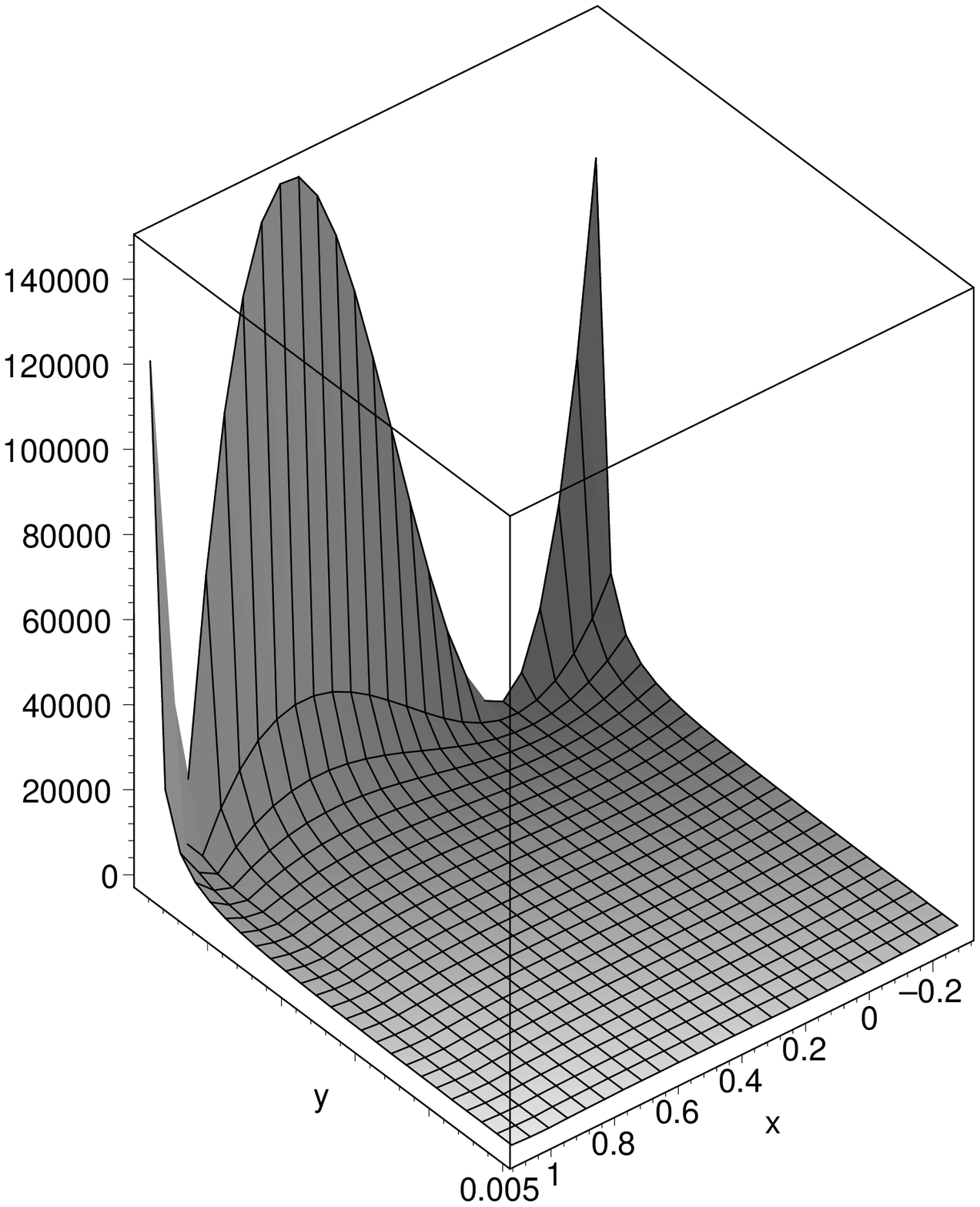, width=.4\textwidth}{\label{T60011}Contour plot of the
  potential generated by a brane wrapping the $q_0 = 0$, $q = 0$, $p =
  1/3$, $p^0 = -1$ cycle in a six dimensional factorizable torus with
  the same complex structure in the three 2-tori.}
  {\label{T600113D}Three dimensional representation of the
  potential generated by a brane wrapping the $q_0 = 0$, $q = 0$, $p =
  1/3$, $p^0 = -1$ cycle in a six dimensional factorizable torus with
  the same complex structure in the three 2-tori.}

\

a.2.iii) if $\Delta < 0$, there is one real root and two complex
conjugates. The minimum is in the interior of the complex structure
moduli. The potential goes to zero at that point. See Figs.
\ref{T61001} and \ref{T610013D}. Note that this possibility cannot be
achieved with a factorizable cycle. Following Moore we can suspect
that the BPS state does not exist. One interesting case when precisely
this happens is if we take the combination of two factorizable cycles:
$(1,0)(1,0)(1,0)$ and $(0,1)(0,1)(0,1)$. It is easy to check that the
minimum is when the two states do not form a bound state. The minimum
is at $\tau = i$, where the two branes have angles $\theta_i = \pi/2$,
i.e. at the centre of the tetrahedron defined by the masses of the
scalars that can become tachyons, see Fig.~\ref{T6}. They cannot decay
into a bound state.

\smallskip \DOUBLEFIGURE[b]{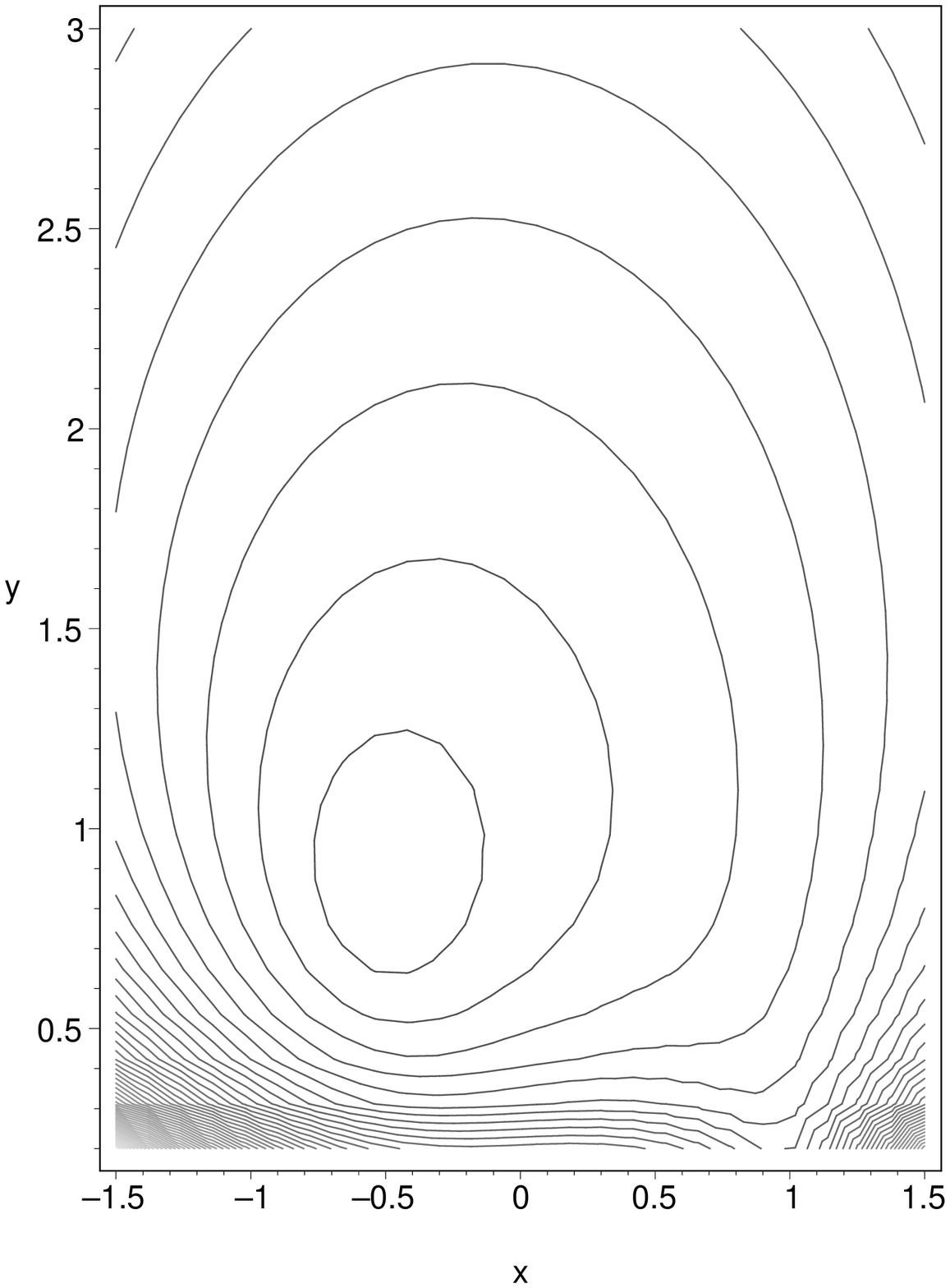, width=.4\textwidth}
{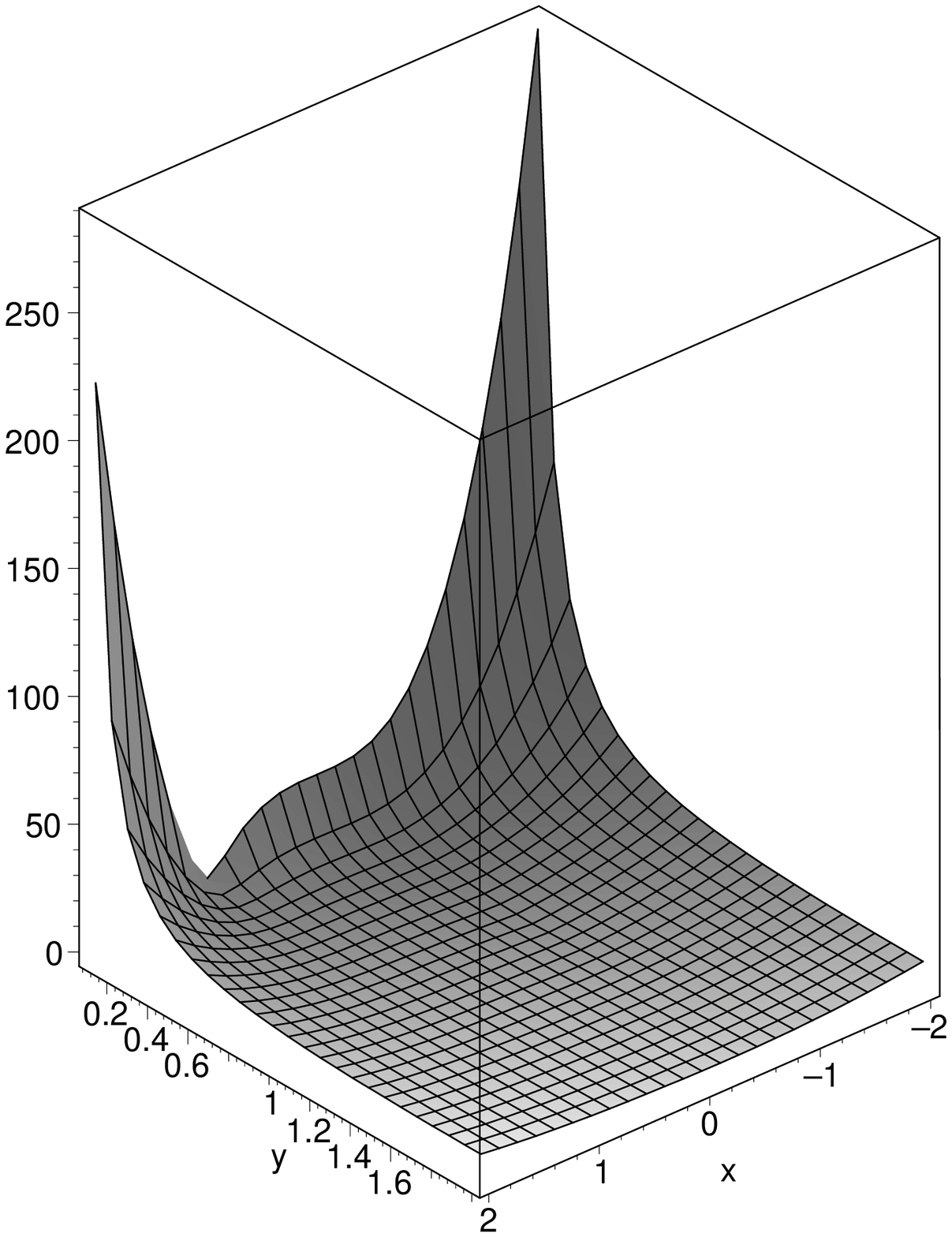, width=.4\textwidth}{\label{T61001}Contour plot of the
  potential generated by a brane wrapping the $q_0 = 1$, $q = 0$, $p =
  0$, $p^0 = -1$ cycle in a six dimensional factorizable torus with
  the same complex structure in the three 2-tori.}
  {\label{T610013D}Three dimensional representation of the
  potential generated by a brane wrapping the $q_0 = 1$, $q = 0$, $p =
  0$, $p^0 = -1$ cycle in a six dimensional factorizable torus with
  the same complex structure in the three 2-tori.}

\

\noindent
b) If the metric is factorisable in two tori, one 4-dimensional, the
other 2-dimensional, $T^6=T^4\times T^2$, we recover the previous
lower-dimensional cases, and the system will be driven to the
boundary.  An specific example of this behaviour is to consider that
the 3-cycles are factorised into 2-cycles wrapping the 4-dimensional
torus and only 1-cycle in the 2-dimensional torus. Then, from the
general analysis to be discussed below, one can see that $\Delta=0$.

\noindent 
c) If the metric cannot be factorised. In this case we have
to study the general solution, as described in Ref.~\cite{Moore}. As
we have seen above, the central charge can be taken to be in this case
\begin{equation}
\int_{\Gamma} \Omega_0 = q_0 + Q_{ij} \tau^{ij} + 
P^{ij} {\rm Cof}\,\tau_{ij} - p^0{\rm det}\,\tau \,,
\end{equation}
i.e. the period with $\Omega_0$. The equations for the critical points 
(\ref{critical}) become:
\begin{equation}
\begin{array}{rl}
\Im(2\bar{C}) &= p^0 \,,\\
\Im(2\bar{C}\,\tau^{ij}) &= P^{ij} \,,\\
\Im(2\bar{C}\,{\rm Cof}\,\tau_{ij}) &= - Q_{ij} \,,\\
\Im(2\bar{C}\,{\rm det}\,\tau) &= q^0 \,. 
\end{array}
\end{equation}
Note that there are $b_3=20$ equations and $3\times3+1=10$ complex
unkowns. The solution of this system of equations is described in
Ref.~\cite{Moore}.  Defining,
\begin{equation}
\begin{array}{rl}
R &\equiv {\rm Cof}\,P + p^0 Q \,,\\
M &\equiv 2 {\rm det}\,P + (p^0q_0 + {\rm tr}(PQ))p^0 \,, \\
D &\equiv 2[({\rm tr}PQ)^2-{\rm tr}(PQ)^2]-(p^0q_0 + {\rm tr}PQ)^2 + 
4(p^0\,{\rm det} Q - q_0\,{\rm det} P)\,, \\
\end{array}
\end{equation}
the solution exists for ${\rm det}R \neq 0$, and $D>0$. The result for
  a general cycle is given by~\cite{Moore} 
\begin{eqnarray}
\tau &=& {1\over2R}\left[2PQ - (p^0q_0 + {\rm tr}(PQ)) + 
\frac{i}{2}\sqrt{D}\right] \,, \nonumber \\ 
2 \bar{C} & =& \frac{M}{\sqrt{D}} + i p^0\,.
\end{eqnarray}
The value of the potential at the critical point is:
\begin{equation}
V_0(\phi) = e^{-\phi} \sqrt{D} \,.
\end{equation}
There are three different cases:
\begin{itemize}

\item $D > 0$. There is a relation between $p$, $M$ and ${\rm det}R$,
i.e. $4\,{\rm det}R = M^2 + p^2 D$. So in this case $D>0 \Rightarrow
{\rm det}R>0$. There is a solution and the brane exists at the minimum.
  
\item $D = 0$. We are in a boundary of the moduli space, $\Im\,
\tau = 0$.
  
\item $D < 0$. There is no BPS state with these charges in the
minimum. The system will decay into a set of branes.

\end{itemize}

Let us now compare with the factorizable cycles we are familiar
with. Consider generically three 1-cycles $(n_1,m_1)(n_2,m_2)(n_3,m_3)$. 
Then
\begin{equation}
\begin{array}{rl}
Q_{ij} &= {\rm diag}(n_2 n_3 m_1, n_1 n_3 m_2, n_2 n_1 m_3 ) \,, \\
P^{ij} &= {\rm diag}(m_2 m_3 n_1, m_1 m_3 n_2, m_2 m_1 n_3 ) \,,\\
q_0 & =  n_1 n_2 n_3  \,,\\
p^0& = - m_1 m_2 m_3  \,. 
\end{array}
\end{equation}
It is easy to check that in this case, $D = 0$ and ${\rm det}R=0$, so
there is no solution inside the complex structure moduli space, but
only at the boundaries.  This agrees with the previous results that
for factorizable cycles the minimum of the potential is at the
boundary.

Let us now consider the sum of the $(1,0)(1,0)(1,0)$ and
$(0,1)(0,1)(0,1)$ cycles. In this case $q =- p = 1$, and $Q = P
=0$. Then $D = -1$ is a negative number, which indicates that the
bound state will decay into two states. It is easy to prove that for a
pair a factorizable branes $D = -I$, where $I$ is the number of
intersections between the two branes, a topological number. Then we
can say that the bound state of two branes is always unstable and will
decay to a two brane system. If the complex structure is factorizable
one can easily check that this happens when the angles are
$(\pi/2,\pi/2,\pi/2)$, i.e. at the centre of the tetrahedron of
Fig.~\ref{T6}. The proof is easy, applying $SL(2,\Z)^3$
transformations one can take a general two brane factorizable
configuration to $a:(1,0)(1,0)(1,0)$ and
$b:(n_1,m_1)(n_2,m_2)(n_3,m_3)$.  The minimum, as we have said, will
be a two-state system. Then the potential is proportional to the
sum of the norms of the periods on these cycles. If the complex
structure is factorizable, the minimum will be at:
\begin{eqnarray}
m_i\,\Re\,\tau_i + n_i & = & 0 \nonumber \\
\prod_i |m_i|\,\Im\,\tau_i & = & 1\,.
\end{eqnarray}
The angles $\theta_i$ are defined through
$$\tan\,\theta_i = {m_i\,\Im\,\tau_i\over m_i\,\Re\,\tau_i +
n_i},$$ 
such that at the factorizable minimum they all become $\theta_i = \pi/2$. 
The potential at the minimum is precisely $V_0 = 2e^{-\phi} \sqrt{|I|}$.

Note that by adding more factorizable branes we will never recover a
general cycle because $Q_{ij} = P_{ij} = 0$, for $i \neq j$. That is,
factorizable cycles only span {\em diagonal} $Q$ and $P$ matrices.

Another very interesting example is the following: Three
factorizable cycles:\\ $(-1,0)(1,0)(1,0)$, $(0,-1)(0,1)(0,1)$ and
$(1,1)(1,1)(1,1)$ combine into a general cycle: $q_0 = p_0 =0$, $Q = P
= 1$. Following the same procedure, one can see that $D = 3 > 0$, such
that the initial brane configuration decays to the combined system in
the minimum. The minimum has a complex structure $\tau = (-\half + i
{\sqrt{3}\over4})\,\I$. See Fig.~\ref{T60110}, where the
potential is plotted keeping the complex structure diagonal and equal for
the two dimensional tori. The value of the potential at the minimum is,
as expected, $V_0(\phi) = e^{-\phi} \sqrt{3}$.

\smallskip \DOUBLEFIGURE[b]{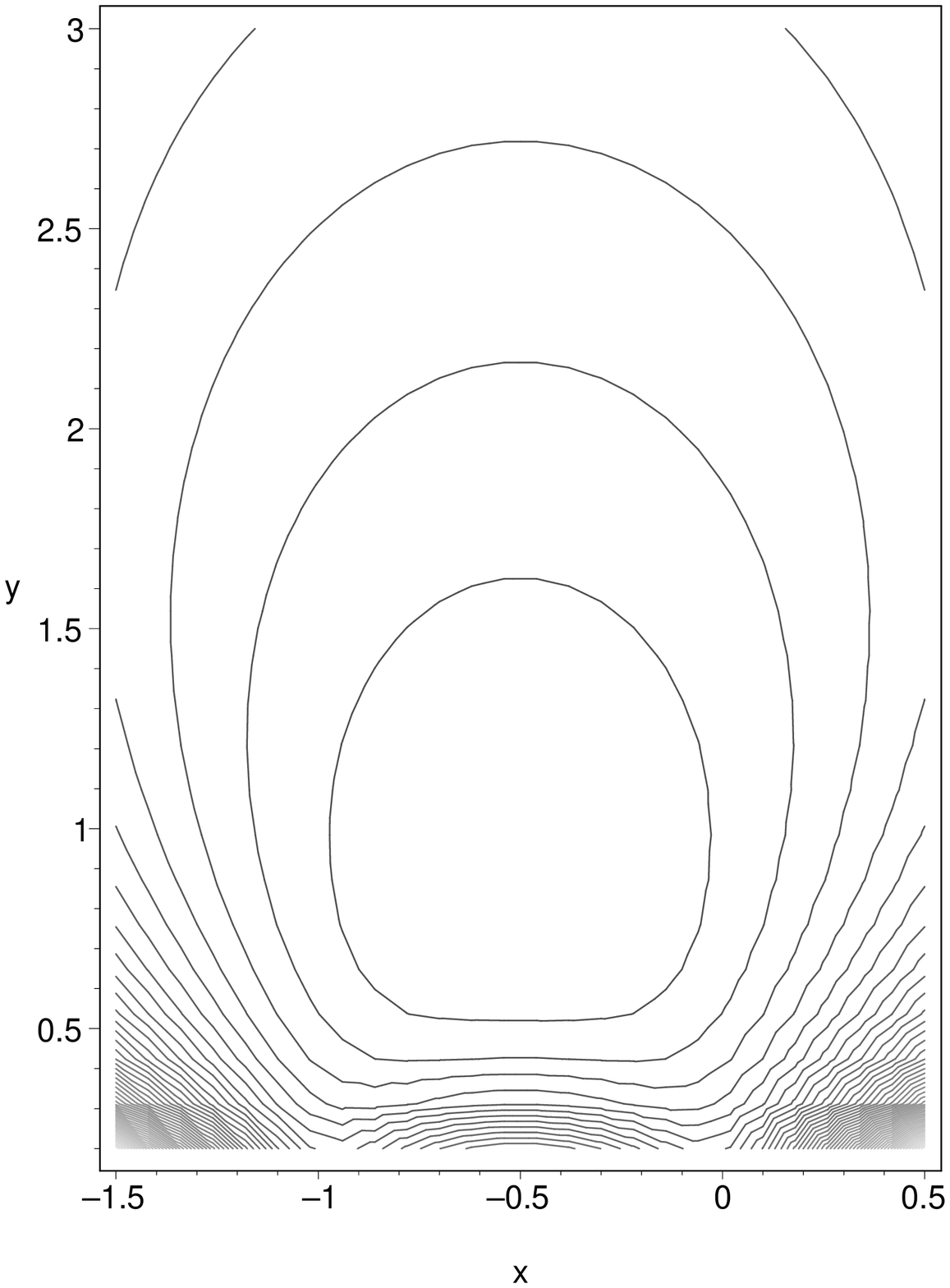, width=.4\textwidth}
{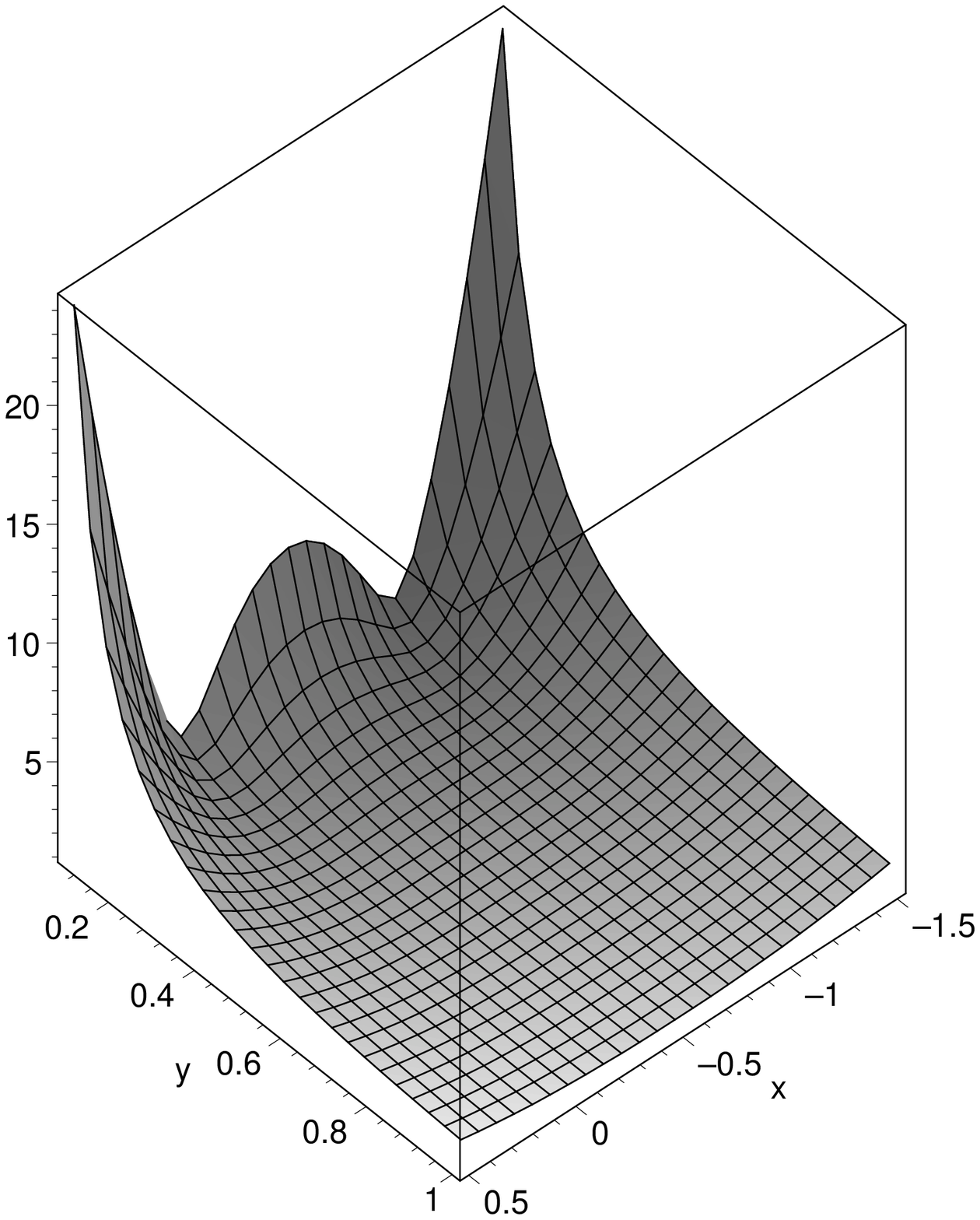, width=.4\textwidth}{\label{T60110}Contour plot of the
  potential generated by a brane wrapping the $q_0 = 0$, $Q = 1$, $P =
  1$, $p^0 = 0$ cycle in a six dimensional factorizable torus with the
  same complex structure in the three 2-tori.}{\label{T601103D}Three
  dimensional representation of the potential generated by a brane
  wrapping the $q_0 = 0$, $Q = 0$, $P = 0$, $p^0 = -1$ cycle in a six
  dimensional factorizable torus with the same complex structure in
  the three 2-tori.}

\section{Stabilising complex structure moduli. Examples.}

In the above examples we have seen different types of behaviours. The
evolution of the complex structure fields can drive them to the
boundary of the moduli space, to a point in the interior of the moduli
space, or can make the brane system to decay by crossing lines of
marginal stability. We will described these three very distinct
behaviours in this section, with specific examples.

\subsection{At the boundary}

The simplest example one can construct with this kind of behaviour is
a brane wrapping a $(0,1)$ cycle in a two dimensional torus. To cancel
the Ramond-Ramond tadpoles one can put an antibrane on the same cycle
but far away from the other in such a way that there is no tachyonic
mode between them. Of course, the one-loop corrections in the open
string description (tree-level in the closed string) will make these
two branes approach one another. However, at large distances it is
sufficient to analyse only the tree-level potential. Within this
aproximation, we find that the effective scalar potential is of
the form
\begin{equation}
V(\phi,\tau) = e^{- \phi} {|\tau |\over\sqrt{\Im\,\tau}} \,.
\end{equation}
The minimum of this potential is at $\tau \rightarrow 0$, i.e. at the
boundary. Since there is one brane that is always minimising the volume,
the D-brane will never decay to another system, but the brane and
antibrane will separate, while the area is kept fixed.

Analogously, a brane wrapping a $(1,0)$ cycle and an antibrane
wrapping the same cycle in the opposite side of the torus will cancel
the R-R tadpoles and produce a potential of the form
\begin{equation}
V(\phi,\tau) = e^{- \phi} \frac{1}{\sqrt{\Im\,\tau}} \,.
\end{equation}
Minimisation of this potential drives the brane to $\Im\tau\rightarrow
\infty$, which means that the two branes will separate and the tachyon
will never appear.

This is a very interesting behaviour that contrasts with the one-loop
correction responsible for the interaction between the two branes. By
adding this interaction, we find two competing effects: NS-NS tadpoles
will take branes far apart from eachother, while the D-brane
interaction will bring them closer and closer. There is a limiting
case in which the two D-branes are just in opposite places in the
compact space. The one-loop effect is vanishing (it is a critical but
unstable point) and the two branes will separate, never decaying into
the vacuum. Alternatively, one can imagine the branes at a distance
such that the two effects compensate eachother: the NS-NS tadpole
potential, at tree-level, being momentarily cancelled by the one-loop
interaction. An interesting physical application of this unstable
equilibrium is precisely that which may drive a relatively long period
of inflation~\cite{inflation}.

\subsection{At a point in the interior}

The simplest system with this kind of behaviour is a six-dimensional
torus with a bound state of two D6-branes, $(1,0)(1,0)(1,0)$ and
$(0,1)(0,1)(0,1)$.  As we have seen in the previous section, the
minimum is at the interior of the complex structure moduli space,
where the bound state has decayed to the two-brane system. This
system is T-dual to a D9 D3-brane system.

Another system, described above, is the bound state of D6-branes:
$(-1,0)(1,0)(1,0)$, $(0,-1)(0,1)(0,1)$ and $(1,1)(1,1)(1,1)$. In this
case the bound state is stable in the minimum of the potential. Of
course, this system does not satisfy the Ramond-Ramond tadpole
conditions. However, we can always put an antibrane wrapped on the
same cycle, but far away in the compact space, as we have already
discussed above.

\section{Conclusions and applications}

In this paper we have applied some previous results of
Moore~\cite{Moore}, derived in the context of BPS quantum black holes,
to the analysis of stability of the critical points of the scalar
potential due to the NS-NS tadpoles in the context of
non-supersymmetric toroidal compactifications, when supersymmetry is
broken by the presence of the D-branes. By studying the structure of
the potential for some set of branes we have found that the minima can
be located at the boundary or at a point in the interior of the
complex structure moduli space. Yet another possibility is that, in
the evolution to the minimum, the system decay to another one, across
lines of marginal stability.

As we have seen in the last section, sometimes the minimum of the
potential is not in the vacuum for Type II strings as one would
expect, but at a point where the non-supersymmetric spectra
decouple. This is analogous to a system of D-branes located at
far away points in the compactified space, as in the example
mentioned in the introduction. NS-NS tadpoles induce a potential that
drives the system to the decompactification limit. That is the usual
runaway behaviour for non-supersymmetric compactifications.

It is also interesting to analyse how the flow is corrected by higher
loop effects. For instance, the interaction between two branes due to
the exchange of closed string modes is a one-loop effect (in the open
string description) and can change drastically the behaviour of the
system. One can imagine some points where the attraction of a
brane-antibrane system (a one loop effect) is compensated by the disk
potential (the NS-NS tadpole). This competing effects can have very
interesting applications for cosmological scenarios, see for instance
Refs.~\cite{inflation}.

Some studies for factorizable cycles and metric have been carried out
recently for the Type 0' in Ref.~\cite{bkl02}, where the system seems
to be driven to a point in the interior of the complex structure
moduli space and for Type I string theory in Ref.~\cite{bklo01}. It
would be very interesting to analyse the general structure of the
minima, i.e. for non-factorisable cycles and metrics, within the
context of non-supersymmetric strings and also for the Type I, where
some complex moduli fields are projected out by the orientifold
projection.

\section*{Acknowledgements}

We would like to thank Fredy Zamora for very enjoyable discussions,
and Fernando Quevedo for useful comments on the manuscript. The work
of JGB is supported in part by a Spanish MEC Fellowship and by CICYT
project FPA-2000-980.

%\listoftables           % ONLY IN DRAFT MODE
%\listoffigures          % ONLY IN DRAFT MODE

\end{document}